\documentclass{aa}  

\usepackage{graphicx}
\usepackage{txfonts}
\def\gapprox{\;\rlap{\lower 2.5pt            
 \hbox{$\sim$}}\raise 1.5pt\hbox{$>$}\;}       
\def\lapprox{\;\rlap{\lower 2.5pt            
 \hbox{$\sim$}}\raise 1.5pt\hbox{$<$}\;} 
\begin{document}
   \title{Accuracy requirements to test the applicability of the random cascade model to supersonic turbulence}
   \titlerunning{Accuracy requirements for testing the random cascade model}
   \subtitle{}
   \author{Doris Folini\inst{1} \and  Rolf Walder\inst{1}}
   \institute{\'{E}cole Normale Sup\'{e}rieure, Lyon, CRAL, UMR CNRS 5574, 
           Universit\'{e} de Lyon, France\\
           \email{doris.folini@ens-lyon.fr}
           }
   \date{Received ... ; accepted ...}
  \abstract
  {A model, which is widely used for inertial rang statistics of
    supersonic turbulence in the context of molecular clouds and star
    formation, expresses (measurable) relative scaling exponents
    $Z_{p}$ of two-point velocity statistics as a function of two
    parameters, $\beta$ and $\Delta$. The model relates them to the
    dimension $D$ of the most dissipative structures,
    $D=3-\Delta/(1-\beta)$. While this description has proved most
    successful for incompressible turbulence ($\beta=\Delta=2/3$, and
    $D=1$), its applicability in the highly compressible regime
    remains debated. For this regime, theoretical arguments suggest
    $D=2$ and $\Delta=2/3,$ or $\Delta=1$.  Best estimates based on 3D
    periodic box simulations of supersonic isothermal turbulence yield
    $\Delta=0.71$ and $D=1.9$, with uncertainty ranges of $\Delta \in
    [0.67, 0.78]$ and $D \in [2.04,1.60]$. With these 5-10\%
    uncertainty ranges just marginally including the theoretical
    values of $\Delta=2/3$ and $D=2$, doubts remain whether the model
    indeed applies and, if it applies, for what values of $\beta$ and
    $\Delta$.  We use a Monte Carlo approach to mimic actual
    simulation data and examine what factors are most relevant for the
    fit quality.  We estimate that 0.1\% (0.05\%) accurate $Z_{p}$,
    with $p=1,.\,.\,.\,,5$, should allow for 2\% (1\%) accurate estimates of
    $\beta$ and $\Delta$ in the highly compressible regime, but not in
    the mildly compressible regime. We argue that simulation-based
    $Z_{p}$ with such accuracy are within reach of today's computer
    resources. If this kind of data does not allow for the expected
    high quality fit of $\beta$ and $\Delta$, then this may indicate
    the inapplicability of the model for the simulation data. In fact,
    other models than the one we examine here have been suggested.}

   \maketitle
%

%
%
\section{Introduction}
\label{sec:intro}
Supersonic turbulence is a key ingredient in various astrophysical
contexts, from gamma ray bursts~\citep{lazar-et-al:09,
  narayan-kumar:09} or stellar accretion~\citep{walder-folini:08,
  hobbs-et-al:11} to molecular clouds and star
formation~\citep{chabrier-hennebelle:11, federrath-klessen:12,
  padoan-et-al:12, kritsuk-et-al2:13}. A key question is whether this
turbulence, like incompressible turbulence, is characterized by
universal statistics. Results from 3D periodic box simulations of
driven, isothermal, supersonic turbulence~\citep{kritsuk-et-al:07,
  schmidt-et-al:08, pan-et-al:09} are indeed consistent with the
highly compressible variant~\citep{boldyrev:02} of the hierarchical
structure model that was put forward by~\citet{she-leveque:94} for
incompressible turbulence and that was further scrutinized
by~\citet{dubrulle:94} and~\citet{she-waymire:95}. This model is
correspondingly popular in astrophysics. It is employed, for example,
in the interpretation of molecular cloud
observations~\citep{gustafsson-et-al:06, hily-blant-et-al:08} or to
derive a theoretical expression for the density distribution in
supersonic turbulence~\citep{boldyrev-et-al1:02}, which enters
theories of the stellar initial mass
function~\citep{hennebelle-chabrier:08}.

Nevertheless, some doubts remain whether the model really applies to
simulation data of supersonic turbulence and, if so, with what
parameter values. The best-fit model parameters that we are aware
of~\citep{pan-et-al:09} still come with a 5-10\% uncertainty range
that is only marginally compatible with theoretically predicted
parameter values (see below). Here we argue that today's computer
resources should allow for 1-2\% accurate parameter fits in the highly
compressible regime, thereby likely settling the issue. Our claim is
based on a Monte Carlo approach to mimic actual simulation data.

The hierarchical structure model predicts the ratios $Z_{p}$ of
(observable) structure function scaling exponents $\zeta_{p}$,
$p=1,2,3...$ etc., of a 3D velocity field $\mathbf{u}$ as
\begin{equation}
Z_{p} = \frac{\zeta_{p}}{\zeta_{3}} =
(1 - \Delta)\frac{p}{3} + \frac{\Delta}{1-\beta}\left( 1 - \beta^{p/3} \right).
\label{eq:dubrulle}
\end{equation}
Here, $D=3-C$ is the dimension of the most intermittent structure,
$C=\Delta/(1-\beta)$ the associated co-dimension, $\beta \in
[0,1]$ measures the intermittency of the energy cascade, and
$\Delta \in [0,1]$ measures the divergent scale dependence of
the most intermittent structures. The $\zeta_{p}$ are defined in the inertial range by
\begin{equation}
S_{p}(r) \equiv < | \mathbf{u}(\mathbf{x} + \mathbf{r}) - \mathbf{u}(\mathbf{x})  |^{p} > 
         \,\,\, \propto r^{\zeta_{p}}, 
\label{eq:struc_func_def}
\end{equation}
where $<\dots>$ denotes the average over all positions $\mathbf{x}$
within the sample and over all directed distances $\mathbf{r}$. The
$Z_{p}$ should be well defined over a larger range because of extended
self-similarity~\citep{benzi-et-al:93} and Eq.~\ref{eq:dubrulle}
should remain formally valid for generalized structure
functions $\tilde{S}_{p}(r)$, computed from mass-weighted velocities
$\mathbf{v} \equiv \rho^{1/3}\mathbf{u}$~\citep{kritsuk-et-al:07,
  kritsuk-et-al2:07}.

Several special cases of the model that differ in their parameter
values exist in the literature~\citep[see e.g. the review
by][]{she-zhang:09}. The original model by~\citet{she-leveque:94}
applies most successfully to incompressible turbulence with 1D vortex
filaments as most dissipative structures ($D=1$) and parameter values
$\beta = \Delta = 2/3$. For highly compressible turbulence, parameter
values remain debated.  \citet{boldyrev:02} argues that the most
dissipative structures are 2D shocks, thus $D=2$, and chose to keep
$\Delta=2/3$ and set $\beta=1/3$. By contrast,
\citet{schmidt-et-al:08} argue that $\Delta=1$ (implying $\beta=0$)
to be consistent with Burgers turbulence. A few studies used 3D
simulation data, derived sets of $Z_{p}$, and attempted simultaneous
fits of $\beta$ and $\Delta$~\citep{kritsuk-et-al2:07,
  schmidt-et-al:08,schmidt-et-al:09, folini-et-al:14}. The results
  are inconclusive in that fits of similar quality are obtained for
widely different $\beta$-$\Delta$-pairs. Also using 3D simulation data
(1024$^{3}$, Mach 6) but working with density-weighted moments of the
dissipation rate, \citet{pan-et-al:09} simultaneously fitted $\Delta$
and $D$ to their data. They find $\Delta \in [0.67,0.78]$ and $D \in
[2.04,1.60]$, with a best estimate of $\Delta=0.71$ and $D=1.9$, thus
$\beta=0.35$. The range for $\Delta$ is not compatible with the
suggested $\Delta=1$ (see above), and also $\Delta=2/3$ lies only at
the lower-most bound of the inferred range. Both $\Delta$ and
$\beta$ may thus deviate from their incompressible values ($\beta =
\Delta = 2/3$) as the Mach number increases, making simultaneous
determination of $\beta$ and $\Delta$ a must.

The present study is motivated by this still inconclusive situation.
We want to better understand what factors (accuracy / order of
$Z_{p}$; mildly versus highly compressible turbulence) are most
relevant for the fit quality and why widely different
$\beta$-$\Delta$-pairs yield fits of similar quality. We use this
insight to formulate quantitative estimates of what is needed to
obtain 1\% accurate estimates of $\beta$ and $\Delta$. We present
results in Sect.~\ref{sec:results}, discuss them in
Sect.~\ref{sec:discussion}, and conclude in Sect.~\ref{sec:conc}.
\section{Results}
\label{sec:results}
We first show that $\beta$ and $\Delta$ can be uniquely determined from an
associated (i.e. computed via Eq.~\ref{eq:dubrulle}) pair
$Z_{p_{\mathrm{1}}}$ and $Z_{p_{\mathrm{2}}}$. We then illustrate how
uncertainties in $Z_{p}$ map onto the $\beta$-$\Delta$-plane. Finally,
we give estimates on how accurate the $Z_{p}$ have to be to achieve a
desired accuracy of $\beta$, $\Delta$, and $C$.
\subsection{$\beta$ and $\Delta$  from exact $Z_{p}$}
\label{sec:exact_zp_2p}
Consider two values $Z_{p_{\mathrm{1}}}$ and $Z_{p_{\mathrm{2}}}$ that
both fulfill Eq.~\ref{eq:dubrulle} for the same values $(\beta,
\Delta)$.  In the following, we show that $(\beta, \Delta)$ can 
unambiguously (uniquely) be recovered from $Z_{p_{\mathrm{1}}}$ and
$Z_{p_{\mathrm{2}}}$.

\begin{figure}[tp]
\centerline{
\includegraphics[width=8.5cm,height=5.5cm]{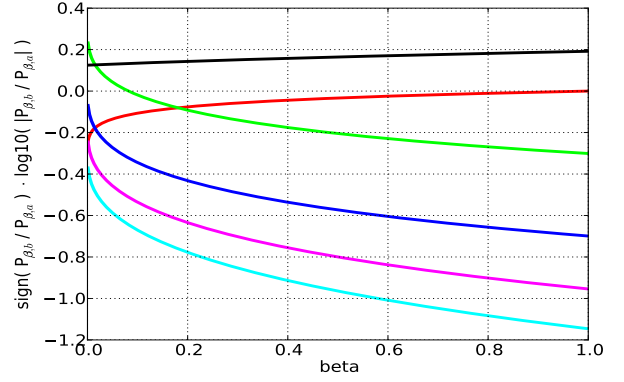}
}
\caption{Ratio of polynomials $P_{\beta,b} / P_{\beta,a}$,
  Eq.~\ref{eq:dubrulle_beta}, (y-axis, shown as logarithm) for
  selected exponents $a$ and $b$ as function of $\beta$ (x-axis).
  Colors indicate $b/a$ = 2 (red), 4 (green), 5 (blue), 6 (magenta), 7
  (cyan), all with $a=p_{\mathrm{1}}/3=1/3$, as well as $b/a$ = 7/6 (black)
  with $a=p_{\mathrm{1}}/3=6/3$.}
\label{fig:poly_ratio}
\end{figure}

We start by rewriting Eq.~\ref{eq:dubrulle}, factoring out $\Delta$:
\begin{equation}
0 = \Delta \left ( \frac{3(1-\beta^{p/3}) - p(1-\beta)}{3(1-\beta)} \right ) +
           \left ( \frac{p - 3 Z_{p_{j}}}{3}\right ), \quad j=1,2.
\label{eq:dubrulle_recast}
\end{equation}
Using $Z_{p_{\mathrm{1}}}$, we can obtain an expression for $\Delta$,
\begin{equation}
\Delta = \left ( \frac{3 Z_{p_{\mathrm{1}}} - p_{\mathrm{1}}}{3} \right ) 
         \left ( \frac{3(1-\beta)}{3(1-\beta^{p_{\mathrm{1}}/3}) - p_{\mathrm{1}}(1-\beta)} \right ).
\label{eq:dubrulle_delta}
\end{equation}
By now writing Eq.~\ref{eq:dubrulle_recast} with
$Z_{p}=Z_{p_{\mathrm{2}}}$, using Eq.~\ref{eq:dubrulle_delta} to
replace $\Delta$, do some re-ordering of terms, and abbreviating
$p_{\mathrm{1}}/3 \equiv a$ and $p_{\mathrm{2}}/3 \equiv b$, we end up
with the following equation for $\beta$:
\begin{equation}
0 = \left ( \frac{-\beta^{b} + b \beta + 1 - b}{-\beta^{a} + a \beta + 1 - a}  \right ) -
    \left ( \frac{Z_{p_{\mathrm{2}}} - b}{Z_{p_{\mathrm{1}}} - a} \right )
  \equiv \frac{P_{\beta,b}}{P_{\beta,a}} - R_{a,b},
\label{eq:dubrulle_beta}
\end{equation}
or
\begin{equation}
R_{a,b} P_{\beta,a} = P_{\beta,b}.
\label{eq:dubrulle_beta_short}
\end{equation}
The polynomial $P_{\beta,x} \equiv -\beta^{x} + x \beta + 1 - x$, with
$x>0$ and $\beta \in (0,1)$, is a monotonically decreasing
(increasing) function for $x<1$ ($x>1$), as can be seen by taking the
derivative of $P_{\beta,x}$ with respect to $\beta$ and as illustrated
in Fig.~\ref{fig:poly_ratio}.  Consequently,
Eq.~\ref{eq:dubrulle_beta_short} has a unique solution, $\beta$, from
which $\Delta$ can be recovered via Eq.~\ref{eq:dubrulle_delta}. Thus
Eq.~\ref{eq:dubrulle} defines an exact one-to-one correspondence
between pairs $(Z_{p_{\mathrm{1}}}, Z_{p_{\mathrm{2}}})$ and $(\beta,
\Delta)$.

Two more points deserve to be highlighted, with the help of
Fig.~\ref{fig:poly_ratio}. The ratio $P_{\beta,b} /
P_{\beta,a} = R_{a,b}$ is shown as a function of $\beta$ for different $a$ and
$b$ or, equivalently, $p_{\mathrm{1}}$ and $p_{\mathrm{2}}$. From the
figure it can be taken that, first, largely different values of
$p_{\mathrm{1}}$ and $p_{\mathrm{2}}$ are advantageous since they result
in stronger stratification of $\beta$ with respect to $R_{a,b} =
(Z_{p_{\mathrm{2}}} - b) / (Z_{p_{\mathrm{1}}} - a)$. The cyan curve
in Fig.~\ref{fig:poly_ratio}, which represents $p_{\mathrm{1}} /
p_{\mathrm{2}} = 1/7$, covers a wider range of values on the y-axis
than the black curve ($p_{\mathrm{1}} / p_{\mathrm{2}} = 6/7$).
Secondly, the stratification is stronger for small $\beta$. Somewhat
anticipating Sect.~\ref{sec:perturbed_zp_2p}, we thus expect
uncertainties in the $Z_{p}$ to be less important if $Z_{p}$ are
available for largely different $p$ and if they are associated with
(yet to be determined) small values of $\beta$.
\subsection{Uncertainty of  $Z_{p}$ in the $\beta$-$\Delta$-plane}
\label{sec:perturbed_zp_2p}
\subsubsection{Single  $Z_{p}$}
\label{sec:zp_in_beta_delta}
\begin{figure}[tbp]
\centerline{
\includegraphics[width=5.1cm]{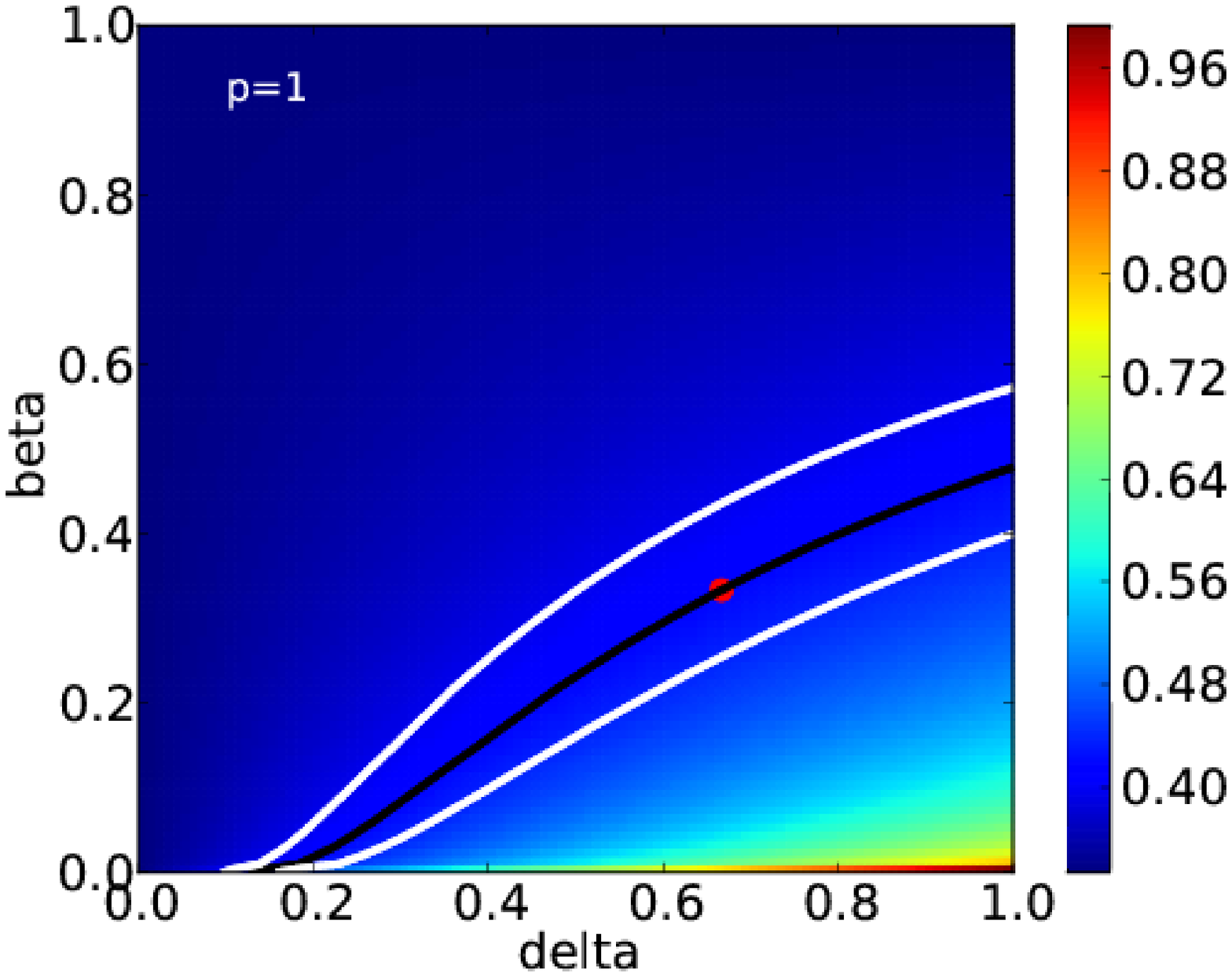}
\hspace{-0.7cm}
\includegraphics[width=5.1cm]{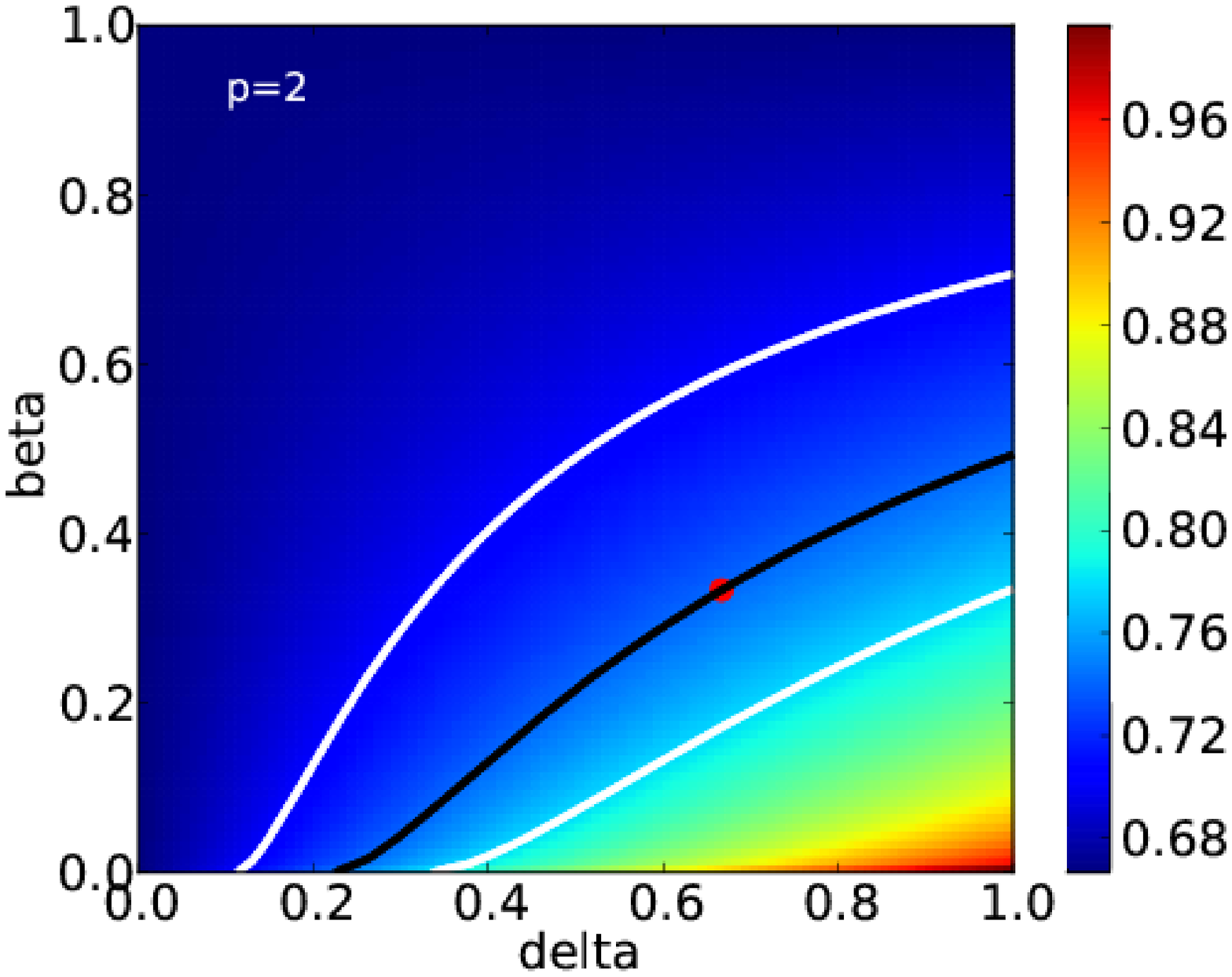}
\hspace{-0.3cm}
}
\centerline{
\includegraphics[width=5.1cm]{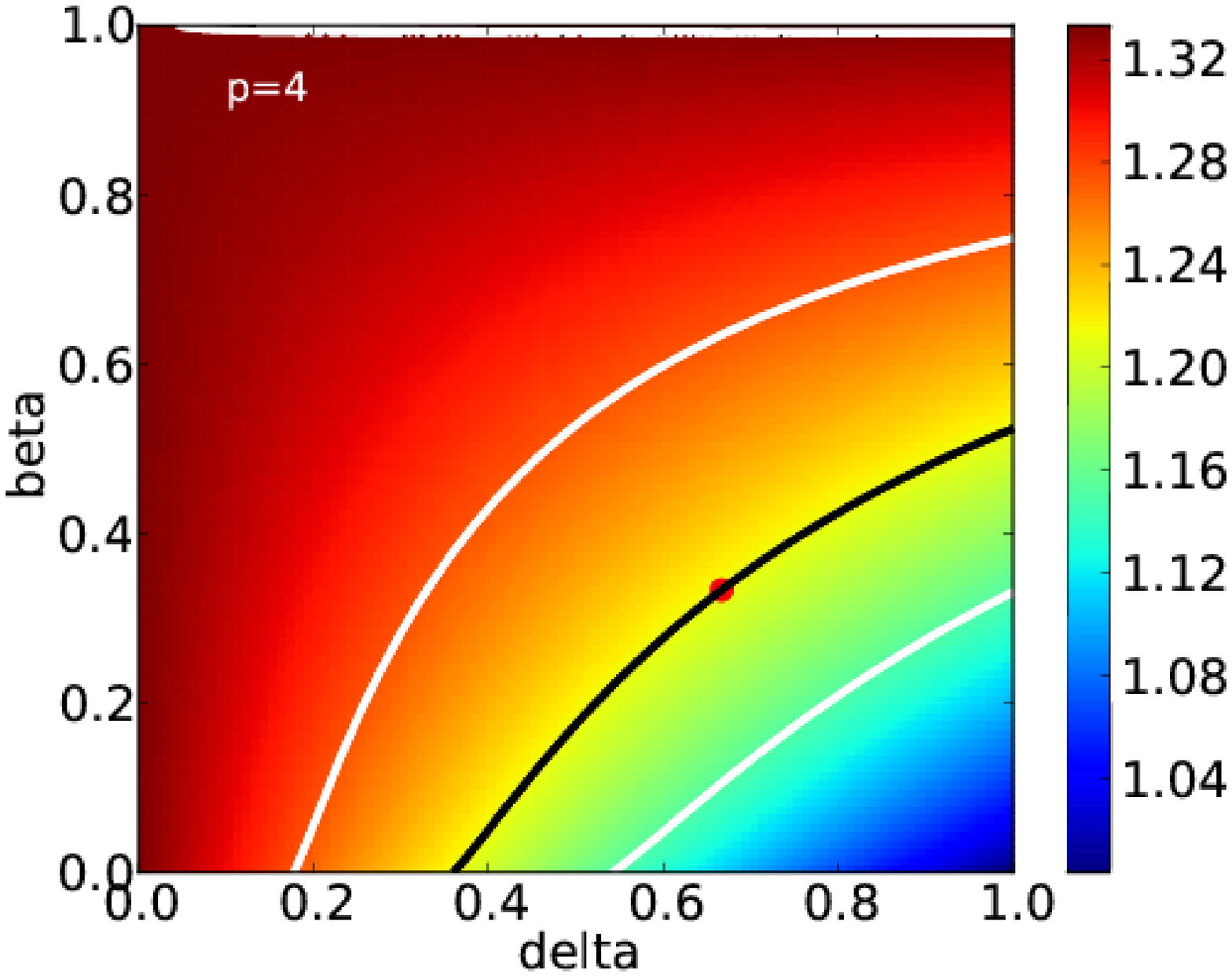}
\hspace{-0.7cm}
\includegraphics[width=5.1cm]{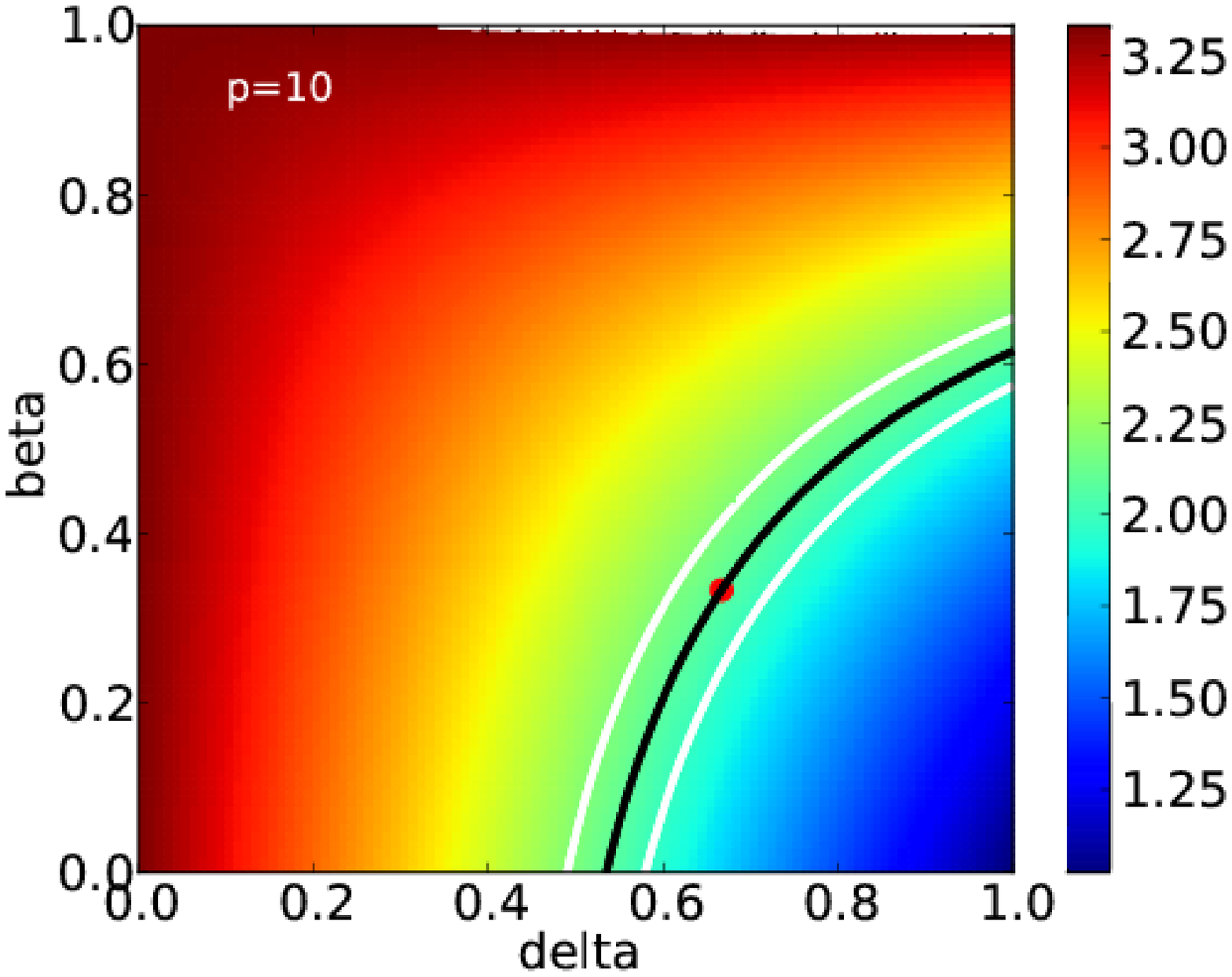}
\hspace{-0.3cm}
}
\caption{$Z_{p}$ in the $\beta$-$\Delta$-plane, role of $p$. Shown is
  $Z_{p}$ (color coded) for $p=1$ (top left), $p=2$ (top right), $p=4$ (bottom left)
  and $p=10$ (bottom right). For $\beta=1/3$ and $\Delta=2/3$ (red dot),
  the curve of constant $Z_{p}$ (black) is shown, as well as curves 
  of  $\pm 5$\% different $Z_{p}$ (white).}
\label{fig:zp_of_beta_delta}
\end{figure}
From Eq.~\ref{eq:dubrulle_delta} it is clear that each $Z_{p}$ defines
a curve in the $\beta$-$\Delta$-plane. If $Z_{p}$ is derived from
model data or observations, it will typically come with an uncertainty
estimate, e.g. $\delta Z_{p}/Z_{p} \le 5$\%, with $\delta$ indicating
the uncertainty. In the $\beta$-$\Delta$-plane, this uncertainty range
translates into an area around the $Z_{p}$ curve. An illustration is
given in Fig.~\ref{fig:zp_of_beta_delta}. The following points may be
made.

One value of $Z_{p}$ (a line of constant $Z_{p}$ in the
$\beta$-$\Delta$-plane) is compatible with a (large) range of $\beta$
and / or $\Delta$ that always includes $\Delta=1$ and $\beta=0$.  The
range tends to be smaller for $Z_{p}$ associated with small $\beta$
and large $\Delta$ (i.e. the lower right corner of
$\beta$-$\Delta$-plane). Uncertainties associated with $Z_{p}$ (5\% in
Fig.~\ref{fig:zp_of_beta_delta}, white curves) augment the range,
especially for $p=2$ and $p=4$, as well as for small $\Delta$ and large
$\beta$ (top left corner of the plane).  Also apparent from
Fig.~\ref{fig:zp_of_beta_delta} (or from taking the derivative with
respect to $\Delta$ of Eq.~\ref{eq:dubrulle}): for fixed $\beta$ and
$p<3$ ($p>3$), $Z_{p}$ is a monotonically increasing (decreasing)
function of $\Delta$. A similar statement holds for $Z_{p}$ as a
function of $\beta$ for fixed $\Delta$.

In summary, we expect uncertainties in the $Z_{p}$ to be more of 
an issue if only low orders of $p$ (up to about 4) are available 
and / or if the (yet to be determined) $\beta$ is large.
\subsubsection{Multiple $Z_{p}$}
\label{sec:multiple_zp}
\begin{figure}[tbp]
\centerline{
\includegraphics[width=5.2cm]{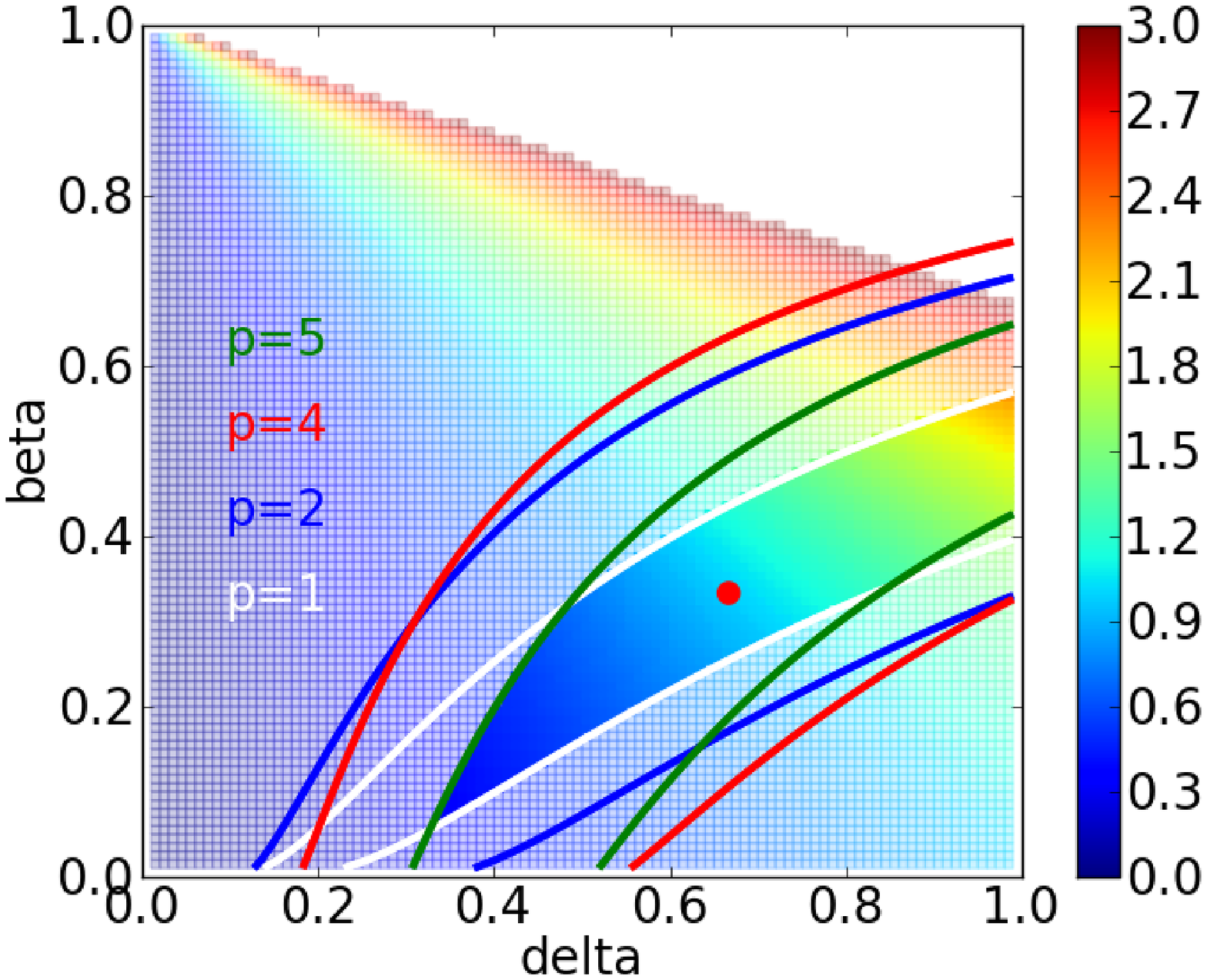}
\hspace{-0.7cm}
\includegraphics[width=5.2cm]{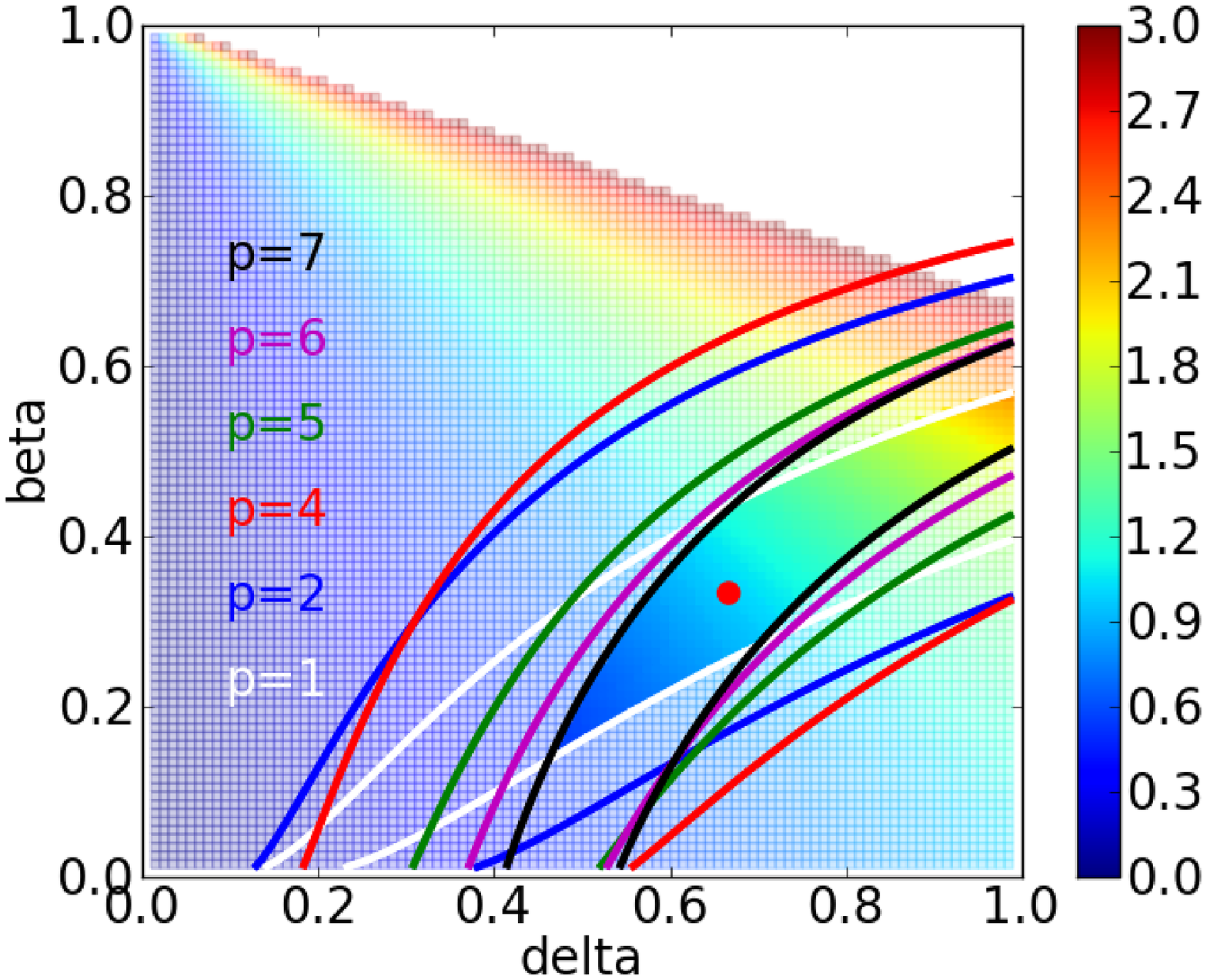}
}
\caption{Part of $\beta$-$\Delta$-plane (color coded in co-dimension
  $C=\Delta / (1-\beta)$, $0<C<3$) within joint reach of (at most) 5\%
  perturbed $Z_{p}$, starting from $Z_{p}$ for $\Delta=2/3$ and
  $\beta=1/3$ (red dot). Left panel: Curves of 5\% perturbed
  $Z_{p}$ values for $p=1$ (white), $p=2$ (blue), $p=4$ (red), and
  $p=5$ (green) and part of the $\beta$-$\Delta$-plane enclosed by all
  of them. Right panel: Same as left panel but also including
  $p=6$ (purple) and $p=7$ (black).}
\label{fig:dmap}
\end{figure}

We now turn to multiple $Z_{p}$ and their associated uncertainty
ranges $\delta Z_{p}$, and ask what area they define in the
$\beta$-$\Delta$-plane. An illustration is given in
Fig.~\ref{fig:dmap}. Starting from one specific pair of $\beta=1/3$
and $\Delta=2/3$ and computing $Z_{p}$ for $p=1,.\,.\,.\,,5$ (left) or $p=1,.\,.\,.\,,7$
(right), we show pairs of 5\% perturbed $Z_{p}$ curves,
i.e. $1.05 \cdot Z_{p}$ and $0.95 \cdot Z_{p}$.

As can be seen, only a small fraction of the $\beta$-$\Delta$-plane
lies between all pairs of perturbed curves. Yet this area comprises a
wide range of $(\beta,\Delta)$ values or co-dimensions.  The 5\%
uncertainty in the $Z_{p}$ translates into a much larger uncertainty
(in per cent) for $\beta$ and $\Delta$. Closer inspection reveals that
the area is actually defined by only two sets of curves: those for
$p=1$ and $p=5$ (left panel) or $p=7$ (right panel). The latter area
is smaller, which indicates that higher order structure functions
constrain the problem of finding $\beta$ and $\Delta$ from a set of
$Z_{p}$ more strongly. Also apparent from Fig.~\ref{fig:dmap} is the
dominant role of the $p=1$ curve for narrowing down the composite area
between all curves.  All this is in line with the expectation (see
Sects.~\ref{sec:exact_zp_2p}) that $Z_{p}$ for largely different $p$
are advantageous for the determination of $\beta$ and $\Delta$.

\begin{figure}[tbp]
\centerline{
\includegraphics[width=5.2cm]{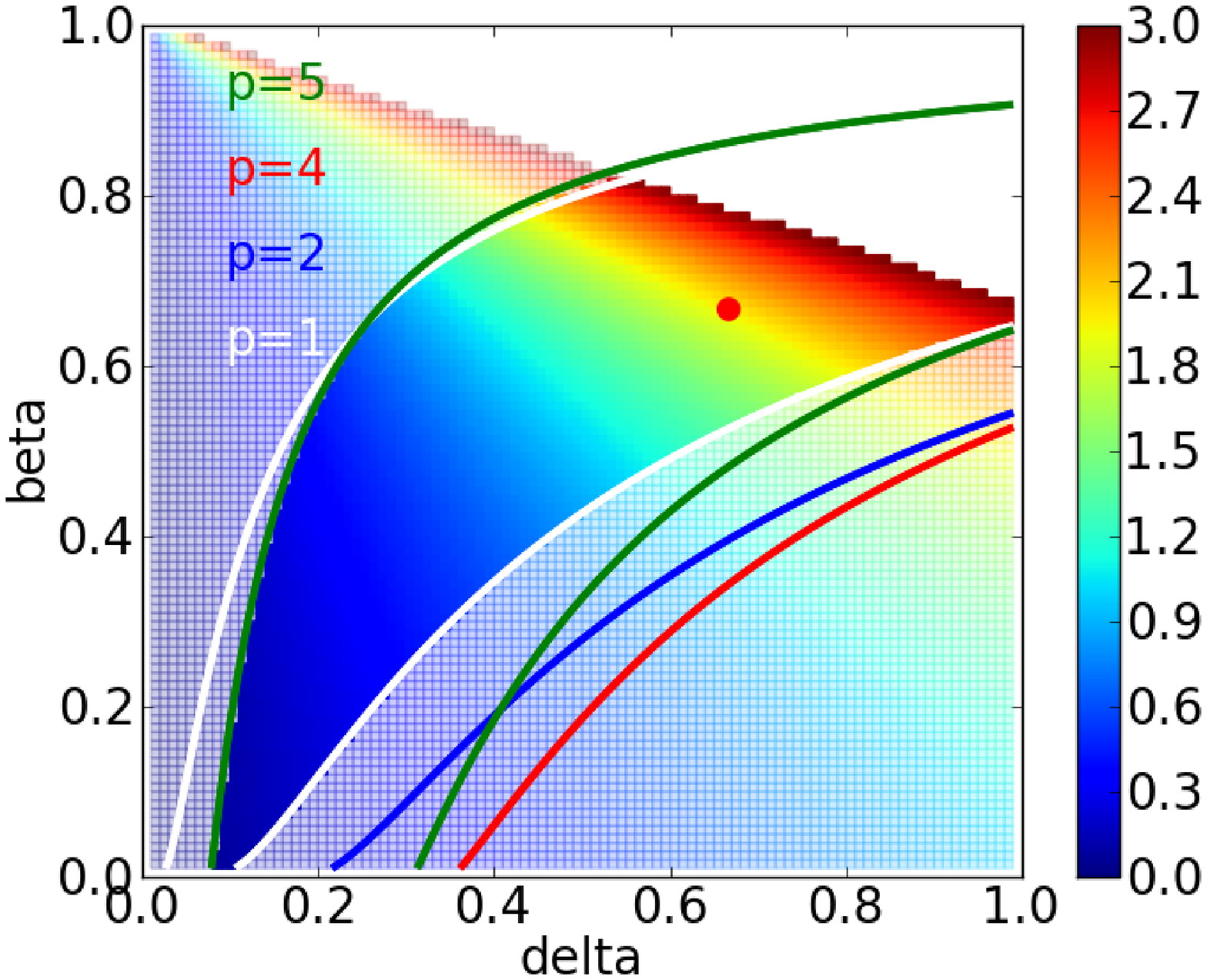}
\hspace{-0.7cm}
\includegraphics[width=5.2cm]{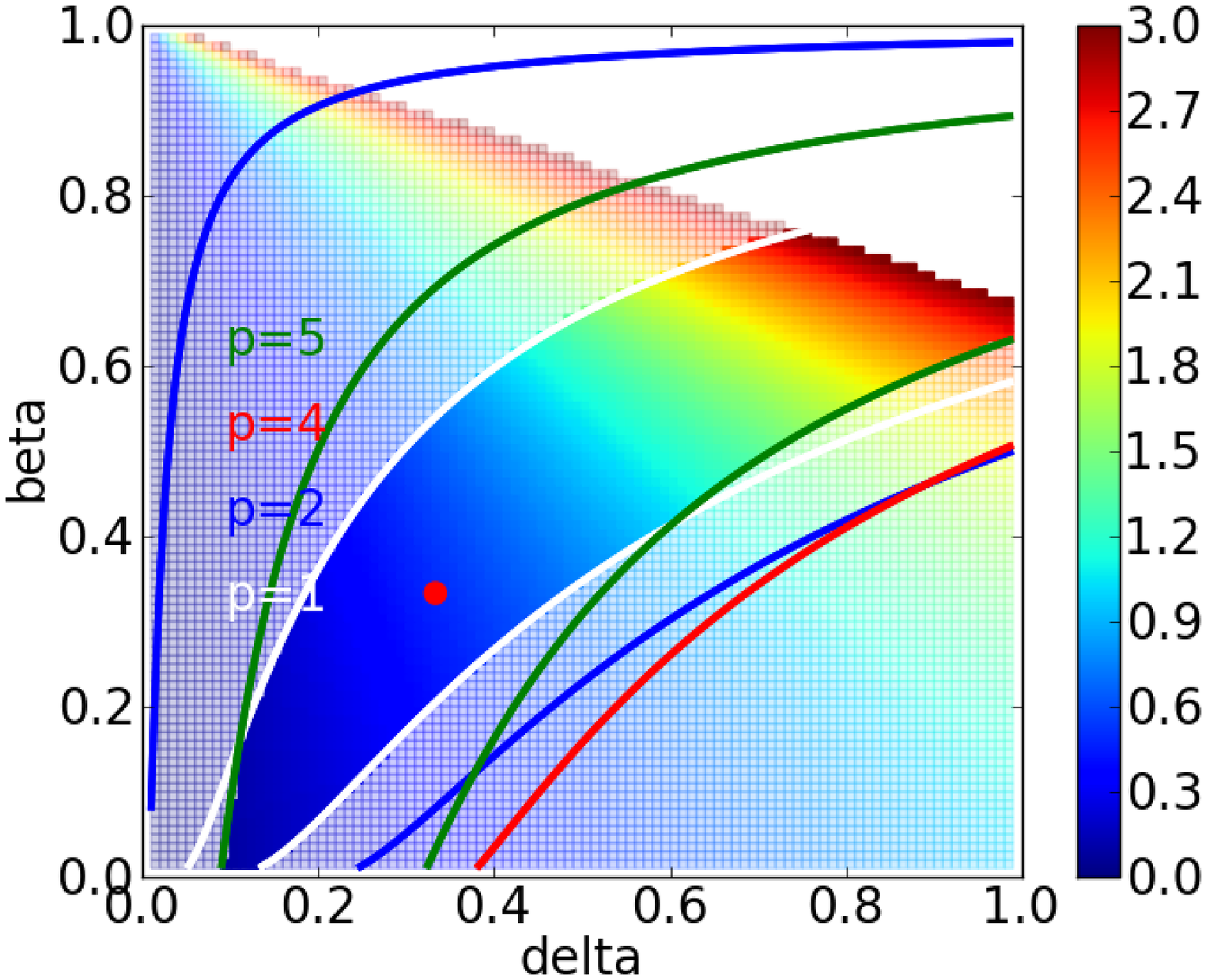}
}
\centerline{
\includegraphics[width=5.2cm]{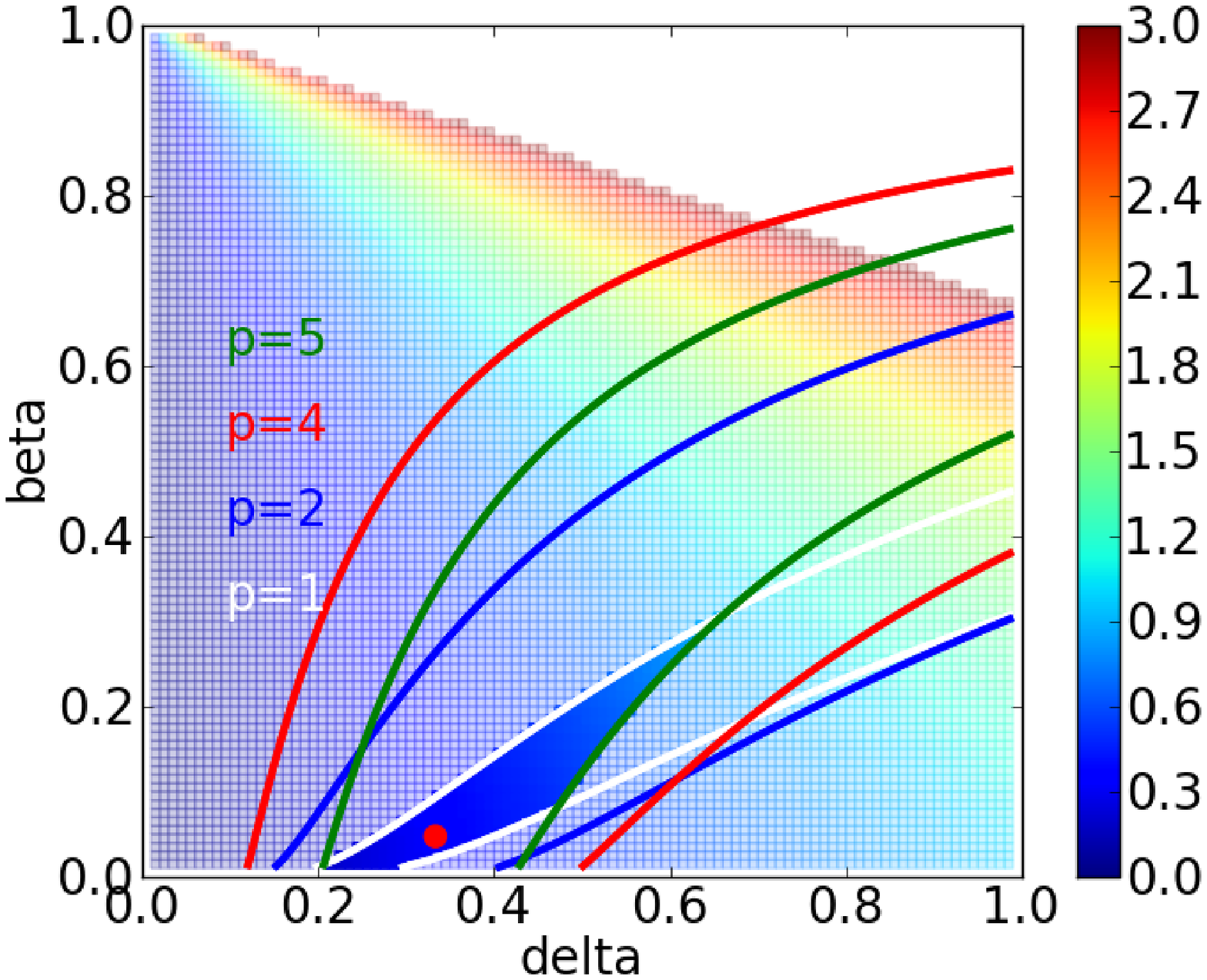}
\hspace{-0.7cm}
\includegraphics[width=5.2cm]{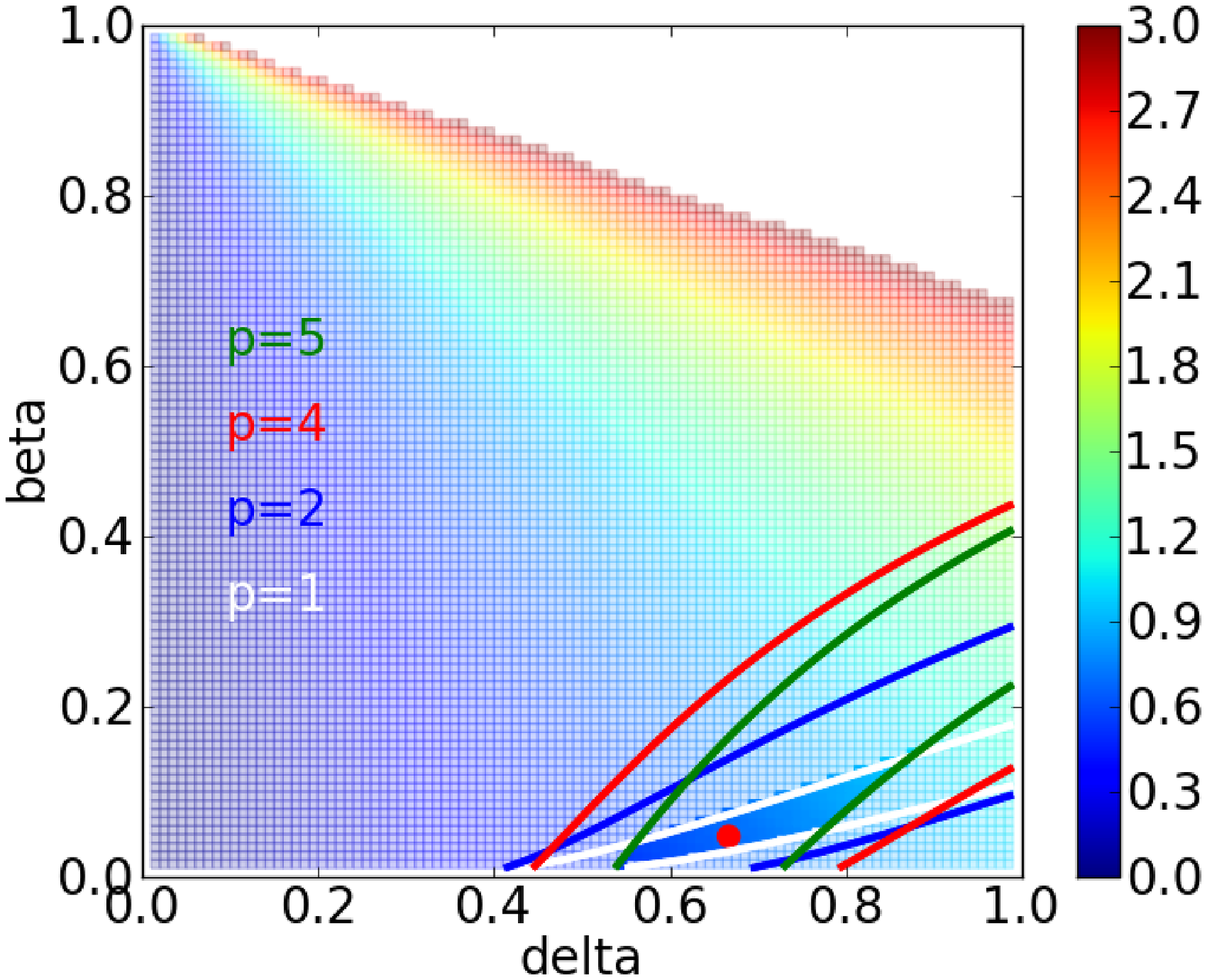}
}
\caption{Same as Fig.~\ref{fig:dmap}, left, but for $\beta=\Delta=2/3$
  (top left), $\beta=\Delta=1/3$ (top right),
  $\beta=0.048$ and $\Delta=1/3$ (bottom left), as well as and
  $\beta=0.048$ and $\Delta=2/3$ (bottom right).}
\label{fig:dmap_var}
\end{figure}

The relevance of the overall location in the $\beta$-$\Delta$-plane is
illustrated in Fig.~\ref{fig:dmap_var}. Again, the area shown is
contained within 5\% perturbed $Z_{p}$ curves for two additional
$(\beta, \Delta)$ pairs.  As can be seen, smaller values of $\beta$
(lower panels) result in smaller areas, independent of $\Delta$. The
crucial role of the $p=1$ (white) and $p=5$ (green) curves for
confining the area persists.  Table~\ref{tab:delta_c} gives a
quantitative idea of the relevance of $\beta$, $\Delta$, $\delta
Z_{p}$, and $p_{\mathrm{max}}$ for the uncertainty range $\pm \delta
C$ of the co-dimensions $C$. A small $\delta C$ basically requires a
small $\delta Z_{p}$, a large $p_{\mathrm{max}}$, a small $\beta$, and
a large $\Delta$. The concrete numbers highlight the difficulty (or
ill-posedness) of the problem. The situation is
worse for larger $\beta$ (bottom rows in Table~\ref{tab:delta_c}) and
better for smaller values of $\beta$ (not shown).

We emphasize that the above considerations serve only as illustration. We
looked at the area confined by a set of $Z_{p} \pm \delta Z_{p}$
curves. We have not yet considered the problem of estimating best-fit
$\beta_{\mathrm{f}}$, $\Delta_{\mathrm{f}}$, and thus $C_{\mathrm{f}}$
for a set of given $Z_{p}$. Such a best-fit solution may lie outside
the area considered here.

In summary, very accurate $Z_{p}$ are needed to derive reliable best
estimates for $\beta$, $\Delta$, and $C,$ and smaller values of $\beta$
help.
\subsection{Best-fit $\beta_{\mathrm{f}}$ and $\Delta_{\mathrm{f}}$
  from uncertain $Z_{p}$}
\label{sec:best_fit_beta_delta}
We now turn to our actual problem of interest: given a set of
perturbed (uncertain) $\tilde{Z}_{p} = Z_{p} + \delta Z_{p}$, what are
associated best-fit estimates for $\beta_{\mathrm{f}}$ and
$\Delta_{\mathrm{f}}$? Different techniques exist to cope with this kind of
question~\citep[e.g.][]{najm:09, lemaitre-knio:10}. We use a simple
Monte Carlo approach.

We start with a pair $(\beta, \Delta)$ and a maximum order
$p_{\mathrm{max}}$, then use Eq.~\ref{eq:dubrulle} to obtain a set of
$Z_{p} = Z_{p} (\beta, \Delta)$ for $p=1$ to $p=p_{\mathrm{max}}$.
Each of these $Z_{p}$ we perturb randomly (uniformly distributed
random numbers) by, at most, $\alpha$\%, which gives us a perturbed set
of $\tilde{Z}_{p}$. For this set of $\tilde{Z}_{p}$ we then seek to
find best-fit $\beta_{\mathrm{f}}$ and $\Delta_{\mathrm{f}}$. In the
following, we do not consider one set of $\tilde{Z}_{p}$, as would be
the case in a real application (unless multiple time slices are
available, see Sect.~\ref{sec:discussion}). Instead, we take a
statistical view for the problem by looking at a large number (1000 to
100\,000, see below) of randomly generated sets of perturbed
$\tilde{Z}_{p}$. This enables us, in a statistical sense, to relate
the accuracy of the $\tilde{Z}_{p}$ with the accuracy of the fitted
parameters. Our approach leaves us with two free parameters, the
uncertainty $\alpha$ and the maximum order $p_{\mathrm{max}}$.

\begin{table}[ht]
  \caption{Illustration of range $\delta C$ of co-dimension $C$ for given order $p$ and 
    uncertainty $\delta Z_{p}$ of structure functions for two $\beta$-$\Delta$ pairs.}
\begin{center}
\begin{tabular}{cc} 
\hline
\multicolumn{2}{c}{$\beta = 1/3$, $\Delta = 2/3$, $C=1$} \\
\rule[-2mm]{0pt}{6mm}      $\delta Z_{p}/Z_{p} = 1$\%    &     $p = 1,.\,.\,.\,,9 $ \\
\hline
$p = 1,.\,.\,.\,,5$:\hspace{0.3cm} $0.79-1.46$ \hspace{0.2cm} & \hspace{0.2cm} $\delta Z_{p}/Z_{p} = 1$\%:\hspace{0.3cm} $0.93-1.08$\\
$p = 1,.\,.\,.\,,6$:\hspace{0.3cm} $0.83-1.23$ \hspace{0.2cm} & \hspace{0.2cm} $\delta Z_{p}/Z_{p} = 2$\%:\hspace{0.3cm} $0.84-1.27$\\
$p = 1,.\,.\,.\,,7$:\hspace{0.3cm} $0.87-1.18$ \hspace{0.2cm} & \hspace{0.2cm} $\delta Z_{p}/Z_{p} = 3$\%:\hspace{0.3cm} $0.74-1.52$\\
$p = 1,.\,.\,.\,,8$:\hspace{0.3cm} $0.90-1.13$ \hspace{0.2cm} & \hspace{0.2cm} $\delta Z_{p}/Z_{p} = 4$\%:\hspace{0.3cm} $0.70-1.92$\\
$p = 1,.\,.\,.\,,9$:\hspace{0.3cm} $0.93-1.08$ \hspace{0.2cm} & \hspace{0.2cm} $\delta Z_{p}/Z_{p} = 5$\%:\hspace{0.3cm} $0.65-2.25$\\
\hline
\multicolumn{2}{c}{$\beta = 2/3$, $\Delta = 2/3$, $C=2$} \\
\rule[-2mm]{0pt}{6mm}      $\delta Z_{p}/Z_{p} = 1$\%    &       $p = 1,.\,.\,.\,,9 $ \\
\hline
$p = 1,.\,.\,.\,,5$:\hspace{0.3cm} $0.51-3.00$ \hspace{0.2cm} & \hspace{0.2cm} $\delta Z_{p}/Z_{p} = 1$\%:\hspace{0.3cm} $1.11-3.00$\\
$p = 1,.\,.\,.\,,6$:\hspace{0.3cm} $0.70-3.00$ \hspace{0.2cm} & \hspace{0.2cm} $\delta Z_{p}/Z_{p} = 2$\%:\hspace{0.3cm} $0.71-3.00$\\
$p = 1,.\,.\,.\,,7$:\hspace{0.3cm} $0.86-3.00$ \hspace{0.2cm} & \hspace{0.2cm} $\delta Z_{p}/Z_{p} = 3$\%:\hspace{0.3cm} $0.55-3.00$\\
$p = 1,.\,.\,.\,,8$:\hspace{0.3cm} $1.00-3.00$ \hspace{0.2cm} & \hspace{0.2cm} $\delta Z_{p}/Z_{p} = 4$\%:\hspace{0.3cm} $0.44-3.00$\\
$p = 1,.\,.\,.\,,9$:\hspace{0.3cm} $1.11-3.00$ \hspace{0.2cm} & \hspace{0.2cm} $\delta Z_{p}/Z_{p} = 5$\%:\hspace{0.3cm} $0.36-3.00$\\
\hline
%
\end{tabular}
\end{center}
\label{tab:delta_c}
\end{table}
\subsubsection{Minimization of least square error in $Z\_{p}$ }
\label{sec:min_least_square}
\begin{figure}[tbp]
\centerline{
\includegraphics[width=4.6cm,height=4.2cm]{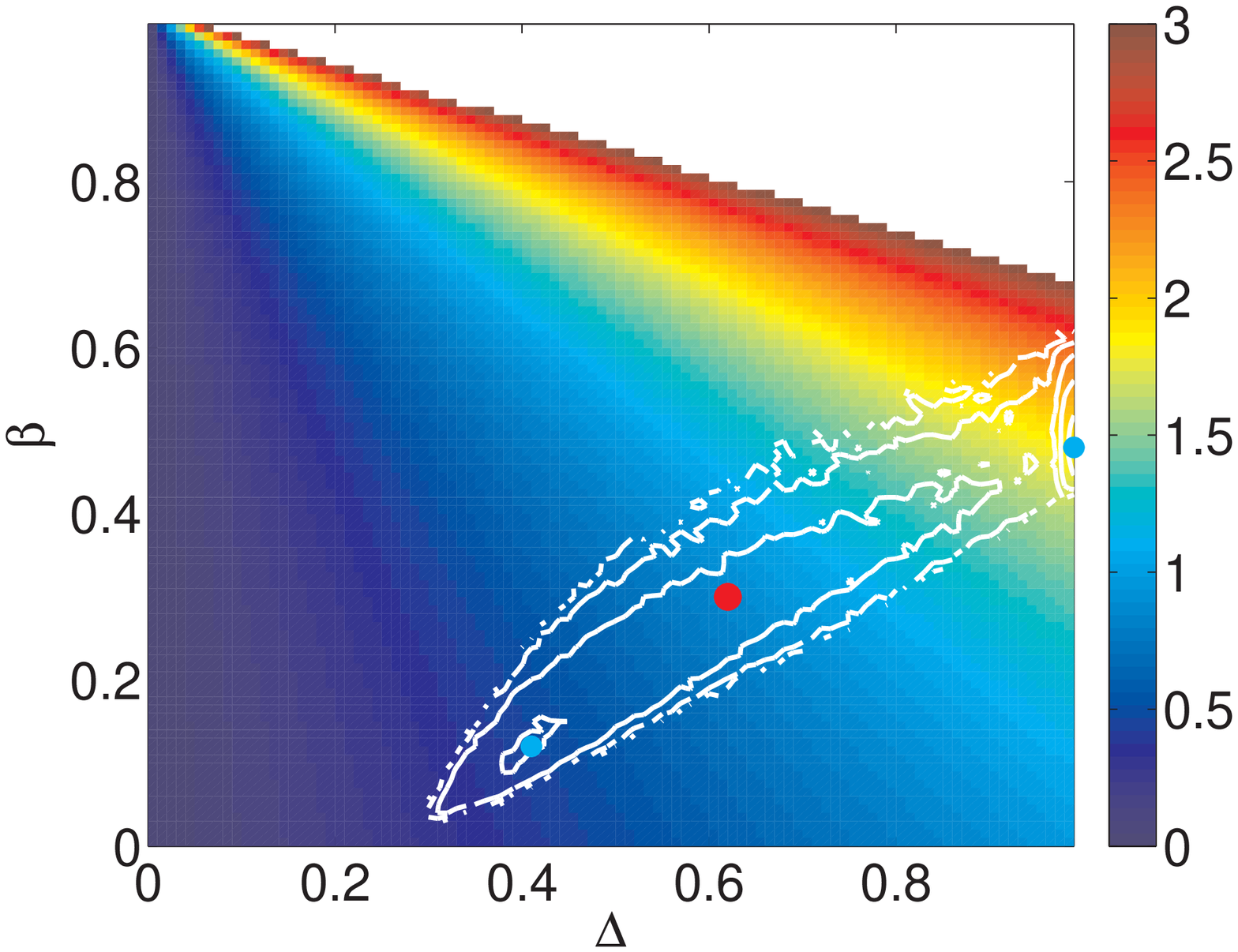}
\includegraphics[width=4.3cm,height=4.2cm]{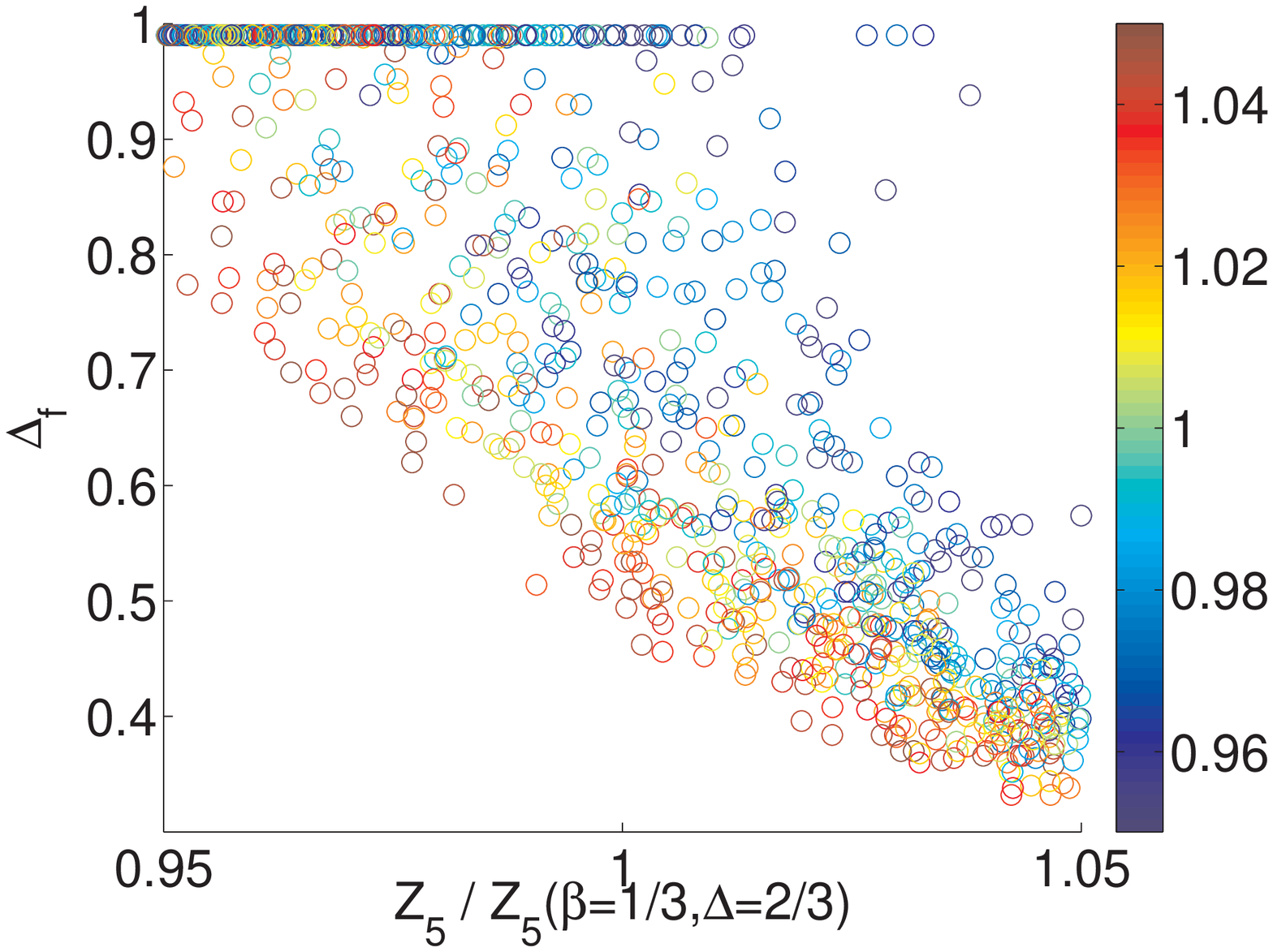}
}
\caption{Best-fit $\beta_{\mathrm{f}}$ and $\Delta_{\mathrm{f}}$ from
  5\% perturbed $Z_{p}$.  Left: 2D histogram (contours, log10,
  spacing 0.5, spanning three orders of magnitude) of best-fit
  $\beta$-$\Delta$-values from 100\,000 perturbed data sets.  We note
  that the 2D histogram shows two peak values (indicated by cyan
  dots), none of them co-located with the unperturbed $(\beta=1/3,
  \Delta=2/3)$ pair (red dot). Right: Underestimation of $Z_{5}$
  favors $\Delta_{\mathrm{f}}=1$. Shown is, for a subset of 1000
  perturbed data sets, $\Delta_{\mathrm{f}}$ as function of
  $\tilde{Z}_{5}/Z_{5}$, again for an unperturbed pair $(\beta=1/3,
  \Delta=2/3)$. Colors indicate $\tilde{Z}_{1}/Z_{1}$.}
\label{fig:stat_map}
\end{figure}
A straightforward way to determine best-fit $\beta_{\mathrm{f}}$ and
$\Delta_{\mathrm{f}}$ for any given set of $\tilde{Z}_{p}$,
$p=p_{\mathrm{min}}.p_{\mathrm{max}}$ is to minimize
\begin{equation}
  \sum_{p=1}^{p_{\mathrm{max}}} 
       [\tilde{Z}_{p} - Z_{p}(\beta_{\mathrm{f}},\Delta_{\mathrm{f}})]^{2}
\label{eq:to_minimize}
\end{equation}
over the $\beta$-$\Delta$-plane. To find the minimum, we compare the
$\tilde{Z}_{p}$ with pre-computed values $Z_{p}(\beta , \Delta)$ on a
fine $\beta$-$\Delta$-grid ($\beta , \Delta \in (0,1)$; grid-spacing
0.002). The associated co-dimension is given by $C_{\mathrm{f}} =
\Delta_{\mathrm{f}} / (1 - \beta_{\mathrm{f}})$. 

To capture the range of potential outcomes for a range of similarly
perturbed data sets $\tilde{Z}_{p}$, we produced 100\,000 perturbed
data sets, for each of which we determined $\beta_{\mathrm{f}}$ and
$\Delta_{\mathrm{f}}$. For initial values $(\beta, \Delta) = (1/3,
2/3)$ and (at most) 5\% perturbed $Z_{p}$ for $p=1,.\,.\,.\,,5$, the result is
summarized in Fig.~\ref{fig:stat_map}.

Shown in the left panel of Fig.~\ref{fig:stat_map} is a 2D histogram
(contours) of our 100\,000 best-fit $(\beta_{\mathrm{f}},
\Delta_{\mathrm{f}})$ pairs. Two points are noteworthy. First, the
overall area defined by the histogram is similar to the area in
Fig.~\ref{fig:dmap}, left panel. This is remarkable since the area in
Fig.~\ref{fig:dmap} is strictly defined by the 5\% uncertainty of the
$Z_{p}$, whereas the area in Fig.~\ref{fig:stat_map} is defined
through a minimization problem. Second, the 2D histogram has an interior
structure with two peaks, around $(\beta, \Delta)=(0.1, 0.4)$ or
$C=0.4$ and $(\beta, \Delta)=(0.45, 1.0)$ or $C=1.8$ (cyan dots). None
of them is co-located with the initial, unperturbed $(\beta, \Delta) =
(1/3, 2/3)$ pair (red dot, $C=1$).

Three questions come to mind. Where do the two peaks in the 2D
histogram come from? Do other $(\beta, \Delta)$ pairs result in a
qualitatively different picture? Can the minimization procedure be
improved to better recover the initial, unperturbed $(\beta, \Delta)$
pair? We address the first two questions in the following while
postponing the third question for Sect.~\ref{sec:alt_best_fits}.

The existence and location of the two peaks can be understood, at
least qualitatively, from two observations.  First, minimization via
Eq.~\ref{eq:to_minimize} gives more weight to larger $p$, as they are
associated with larger values of $Z_{p}$. Roughly speaking, the
best-fit $(\beta_{\mathrm{f}}, \Delta_{\mathrm{f}})$ pair tends to lie
on or close to the curve defined by $\tilde{Z}_{5}$. Moving away from
that curve results in a large penalty in the form of a large
contribution to the sum in Eq.~\ref{eq:to_minimize}. Second, this
translates the minimization problem into the question of where the
curves for $p<5$ come closest to the curve defined by $\tilde{Z}_{5}$.
For illustration, we consider two extreme values of $\tilde{Z}_{5}$.
To stay on the lower green curve in the left panel of
Fig.~\ref{fig:dmap}, ($\tilde{Z}_{5} = 0.95 Z_{5}$) and, at the same
time, be as close as possible to any of the white curves ($p=1$)
results in a $(\beta_{\mathrm{f}}, \Delta_{\mathrm{f}})$ pair to the
right, at $\Delta_{\mathrm{f}}=1$.  By contrast, the upper green curve
(105\% of the exact $Z_{5}$ curve) only comes closest to (intersects)
any white curve between 95\% $Z_{1}$ and 105\% $Z_{1}$ in a region
further to the left. Clearly, the full problem is more intricate, with
also curves for $Z_{2}$ and $Z_{4},$ and the $Z_{5}$ curve not
necessarily adopting one of its two extreme values. Nevertheless,
Fig.~\ref{fig:stat_map}, right panel, suggests the full data to be in
line with the above reasoning. For 1000 randomly picked data
sets from the left panel, we show $\Delta_{\mathrm{f}}$ as a function of
$\tilde{Z}_{5} / Z_{5}$, with $Z_{5}$ the exact value. Colors indicate
$\tilde{Z}_{1} / Z_{1}$.  As can be seen, $\Delta_{\mathrm{f}}=1$
indeed tends to be associated with small $\tilde{Z}_{5}$ and small
$\tilde{Z}_{1}$ (lower green and upper white curve in
Fig.~\ref{fig:dmap}, left panel). Particularly low values of
$\Delta_{\mathrm{f}}$ (e.g. $\Delta_{\mathrm{f}} \approx 0.4$) tend to
occur for large $\tilde{Z}_{5}$ and any $\tilde{Z}_{1}$ (upper green
curve and any white curve in Fig.~\ref{fig:dmap}, left panel).

\begin{figure}[tp]
\centerline{
\includegraphics[width=8.8cm,height=10cm]{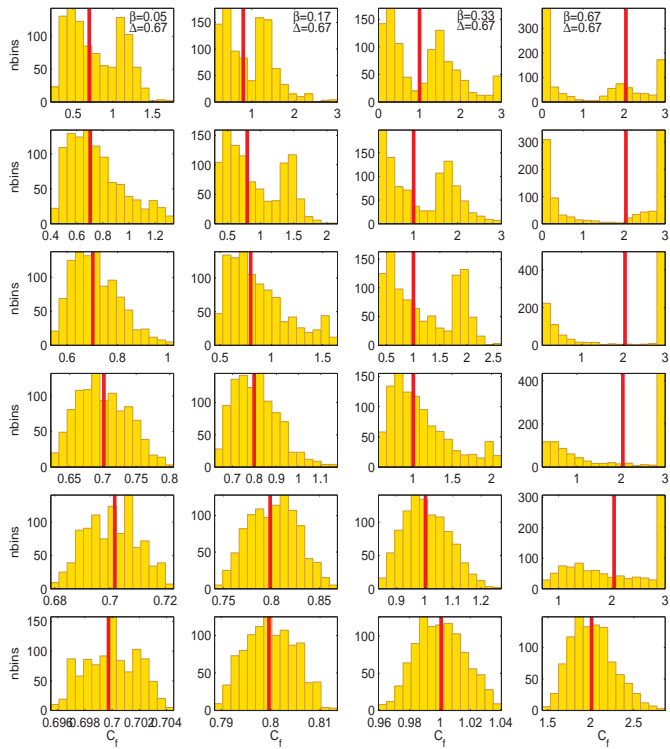}
}
\caption{Role of $\beta$ (columns) and accuracy (rows) of associated
  $Z_{p}$, for fixed $\Delta = 2/3$. Shown are PDFs (y-axis) of
  $C_{\mathrm{f}}$ (x-axis), 1000 random data sets, powers $p=1,.\,.\,.\,,5$
  for $\beta$=0.0476 (first column), $\beta$=0.17 (second
    column), $\beta$=1/3 (third column), and $\beta=2/3$ 
    (fourth column). Corresponding exact co-dimensions (red lines)
  are, from left to right: $C=0.7$, $C=0.8$, $C=1$, and $C=2$.
  Individual rows from top to bottom contain accuracies of 20\%,
  10\%, 5\%, 2\%, 0.5\%, and 0.1\%. As can be seen, the larger $\beta,$
  the more severe are the consequences of inaccuracies in the $Z_{p}$.
  Histograms in the upper right (large $\beta$, low accuracy of the
  $Z_{p}$) look worst.  We note that axis ranges differ among panels,
  to best capture the shape of each histogram.}
\label{fig:accu_vs_codim}
\end{figure}

Concerning other initial values (other exact $(\beta, \Delta)$ pairs),
a similar situation arises in the sense that double peaked histograms
emerge. Details depend, however, on the concrete values of $\beta$ and
$\Delta$, on the assumed uncertainty (5\% or more / less), and on
$p_{\mathrm{max}}$. An illustration is given in
Fig.~\ref{fig:accu_vs_codim}, by means of 1D histograms of
$C_{\mathrm{f}} = \Delta_{\mathrm{f}} / (1 - \beta_{\mathrm{f}})$.
These 1D histograms are less intricate than the 2D histogram in
Fig.~\ref{fig:stat_map}, left, yet still capture the essentials.
We show histograms for $\Delta=2/3$ and different values of $\beta
\in [0.0476, 2/3]$ (corresponding to $C \in [0.7, 2]$) and accuracies
between 0.1\% and 20\%. Five points may be made.  First, the
double-peaked structure that is apparent in the $\beta$-$\Delta$ plane in
Fig.~\ref{fig:stat_map} re-appears as a double peak in the 1D
$C_{\mathrm{f}}$-histograms of Fig.~\ref{fig:accu_vs_codim} (panel in
row three, column three). Second, the double-peak vanishes as $\beta$
and the uncertainty both become small (lower left corner of the
figure). For the same uncertainty, the double-peak exists for large
$\beta$ but not for small $\beta$ (third row in
Fig.~\ref{fig:accu_vs_codim}). For the same $\beta$, the double-peak
exists for large uncertainties but not for small ones (second column
of Fig.~\ref{fig:accu_vs_codim}). Third, going to really small values
of $\beta$ and the uncertainty, the histogram becomes symmetric with
one central peak.  Fourth, only for these really small values is the
co-dimension of the initially prescribed $(\beta, \Delta)$ pair (show
in red) co-located with the peak of the histogram. Fifth, for
$\beta=\Delta=2/3$ (right column) the histogram peaks at
$C_{\mathrm{f}}=3$ instead of $C=2$, unless the accuracy is really
high (0.1\%, last row). This is understandable from the arguments
presented above, with regards to the origin of the two peaks in the 2D
histogram in Fig.~\ref{fig:stat_map}, and from looking at the green
($p=5$) and white ($p=1$) curves in Fig.~\ref{fig:dmap_var}, upper
left panel.

In summary, unless both, $\delta Z_{p}$ and $\beta$ are small, best-fit
values will preferentially reside in either one of the two peaks of
the histograms in Figs.~\ref{fig:stat_map} or~\ref{fig:accu_vs_codim}
instead of merely scattering around the correct solution, as in
Fig.~\ref{fig:accu_vs_codim}, lower left panel.
\subsubsection{Alternative ways to obtain best-fit  $\beta_{\mathrm{f}}$ and $\Delta_{\mathrm{f}}$}
\label{sec:alt_best_fits}
A number of ideas come to mind on how one may improve the best-fit
approach detailed in Sect.~\ref{sec:min_least_square}.

\begin{figure}[tbp]
\centerline{
\includegraphics[width=8.8cm]{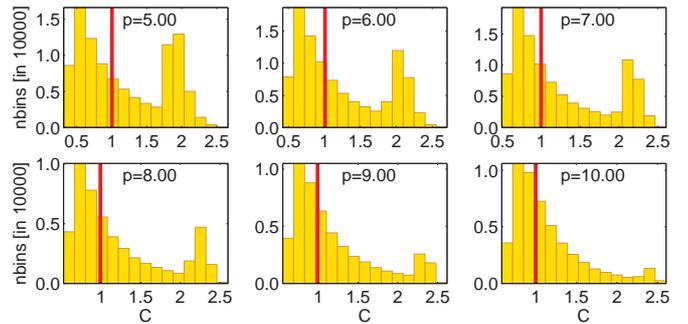}
}
\caption{Role of higher order moments. Inclusion of higher 
  order ESS scaling exponents (from $p=5$, top left, to $p=10$, bottom right)
  gradually reduces the erroneous peak in the best-fit co-dimension
  (around $C_{\mathrm{f}}=2.5$). Shown are PDFs of $C_{\mathrm{f}}$
  for at most 5\% perturbed $Z_{p}$ values (100\,000 random data sets)
  and exact pair $(\beta=1/3, \Delta=2/3)$. Spacing of the
  $\beta$-$\Delta$-grid for fitting is 0.002.}
\label{fig:d_b_of_p}
\end{figure}

Recalling the findings in Sect.~\ref{sec:zp_in_beta_delta}, including
higher values of $p$ in the best-fit estimate should improve the
situation. From Fig.~\ref{fig:d_b_of_p} it can be taken that this is
indeed the case, at least for the example shown ($\beta=1/3$,
$\Delta=2/3$, accuracy of 5\%). However, the improvement may be
regarded as rather modest. Going from $p=5$ to $p=10$, as is
illustrated in the figure, has about the same effect as staying with
$p=5$ but going from an accuracy of 5\% to an accuracy of 2\%.  From a
practical point of view it also seems questionable whether high order
structure functions can meet the accuracy requirements.  In numerical
simulations, higher order structure functions are probably more prone
to the bottleneck effect~\citep{dobler-et-al:03, kritsuk-et-al:07}.

Another way to improve the situation could be to go to weighted root
mean squares instead of the unweighted sum in
Eq.~\ref{eq:to_minimize}. Hopefully this breaks the dominant
role of the highest order $Z_{p}$ available (see
Sect.~\ref{sec:min_least_square}) and, ultimately, leads to more
accurate best-fit $\beta_{\mathrm{f}}$ and $\Delta_{\mathrm{f}}$.  Two
weightings come to mind.  On the one hand, weights proportional to the
inverse of the $Z_{p}$ with the goal of giving equal weight to each
term in the sum, thus reducing the "overweight" of larger $p$ in the
sum.  On the other hand, we could try to give more weight to $p$ terms
with a higher accuracy (smaller $\delta Z_{p}$).  Corresponding
information may be available, e.g. from the numerical determination of
the $Z_{p}$. We tried both ideas but neither choice of weights
decidedly improved the best-fit values. Weighting tends to change the
relative height of the two peaks in the double peaked histograms of
Fig.~\ref{fig:accu_vs_codim}, but it does not get rid of the double
peaked structure.

We interpret this finding in the following way. First, there are
likely always several $Z_{p}$ that do not have their exact values and
thus draw the solution in different directions, away from its exact
value. Second, the different curve shapes are important so that, even for
weighted sums, the terms $p=1$ and $p=p_{\mathrm{max}}$ are of crucial
importance for the overall fit.

In summary, none of the above alternative ways of fitting
simultaneously for $\beta$ and $\Delta$ provides clearly superior
results to what can be obtained from the straightforward minimization
of Eq.~\ref{eq:to_minimize}. We conclude that, for successful two-parameter
fits of $\beta$ and $\Delta$, highly accurate $Z_{p}$ are a must. A
quantitative estimate of "highly accurate" is given in the next
section.
\subsubsection{Required accuracy of $Z_{p}$ for 'good' best-fit  $\beta_{\mathrm{f}}$ and $\Delta_{\mathrm{f}}$}
\label{sec:accu_needed}
We now ask how accurate the $Z_{p}$ have to be in order to reach a
prescribed accuracy of $C_{\mathrm{f}} = \Delta_{\mathrm{f}} /
(1-\beta_{\mathrm{f}})$ via fitting $\beta_{\mathrm{f}}$ and
$\Delta_{\mathrm{f}}$.

We formulate our accuracy goal in terms of only $C_{\mathrm{f}}$, since we
illustrated in Sect.~\ref{sec:min_least_square} that a single
peaked and roughly symmetric distribution of $C_{\mathrm{f}}$ goes
hand in hand with high accuracy, not only of $C_{\mathrm{f}}$ but also
of the underlying two parameter fit, $\beta_{\mathrm{f}}$ and
$\Delta_{\mathrm{f}}$. If the latter is not accurate
enough, a double peaked distribution for $C_{\mathrm{f}}$ results. We
find, as a rule of thumb, a single peak distribution if 2/3 of all
$C_{\mathrm{f}}$ lie within 10\% or better of the exact $C$. We use
Eq.~\ref{eq:to_minimize} for the two parameter fit, as the more
elaborate attempts of Sect.~\ref{sec:alt_best_fits} gave no decidedly
better results. As theoretical arguments suggest $\Delta = 2/3$ or
larger~\citep{dubrulle:94, schmidt-et-al:08}, we concentrate on that
part of the $\beta$-$\Delta$-plane.

In practical terms, we define a grid of exact pairs $(\beta, \Delta)$
via a (nearly) equidistant grid of $C = 0.4,.\,.\,.\,,2$ and $\Delta =
0.7,.\,.\,.\,,0.99$ plus, in addition, $\Delta=2/3$. We equally define some
fixed levels of perturbations: $\delta Z_{p}/Z_{p}$ (in \%) $\in
[0.05, 0.1, 0.2, 0.5, 1, 2, 5, 10, 15, 20]$. For each exact pair and
each perturbation, we created 1000 perturbed sets of $\tilde{Z}_{p}$,
$p=1,.\,.\,.\,,5$ (see Sect.~\ref{sec:best_fit_beta_delta}). Each perturbed set
is fitted via Eq.~\ref{eq:to_minimize}. For each initial
pair ($\beta$, $\Delta$) and for each prescribed $\delta Z_{p} /
Z_{p}$ this yields 1000 fitted pairs $(\beta_{\mathrm{f}}, \Delta_{\mathrm{f}})$
and derived co-dimensions $C_{\mathrm{f}}$. These can, in principle,
be arranged in histograms, as in Fig.~\ref{fig:accu_vs_codim}. 
Finally, we identify the largest $\delta Z_{p}/Z_{p}$ for which 2/3 of
all $C_{\mathrm{f}}$ lie within the demanded accuracy of the exact,
initial $C$.
\begin{figure}[tbp]
\centerline{\includegraphics[width=8.0cm]{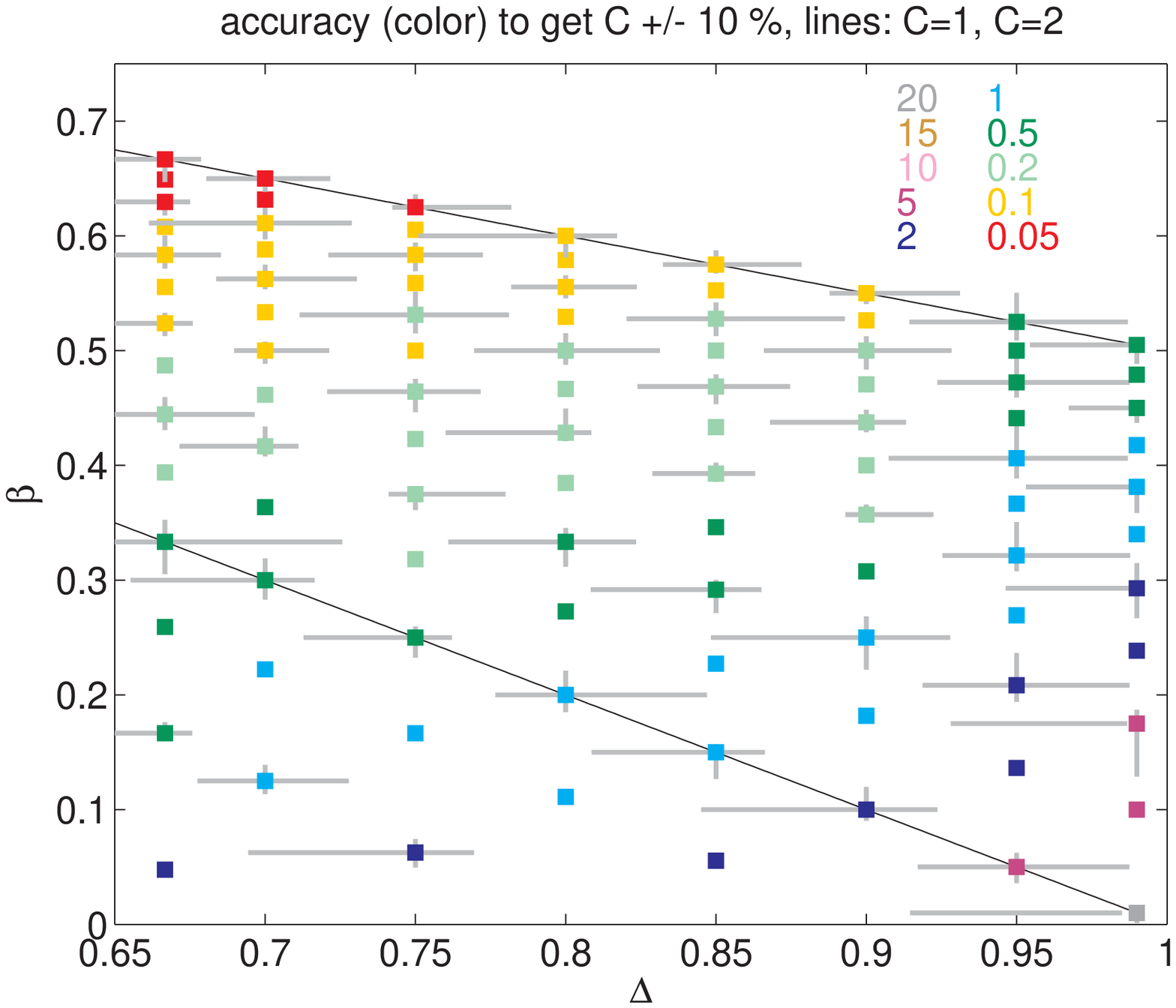}}
\centerline{\includegraphics[width=8.0cm]{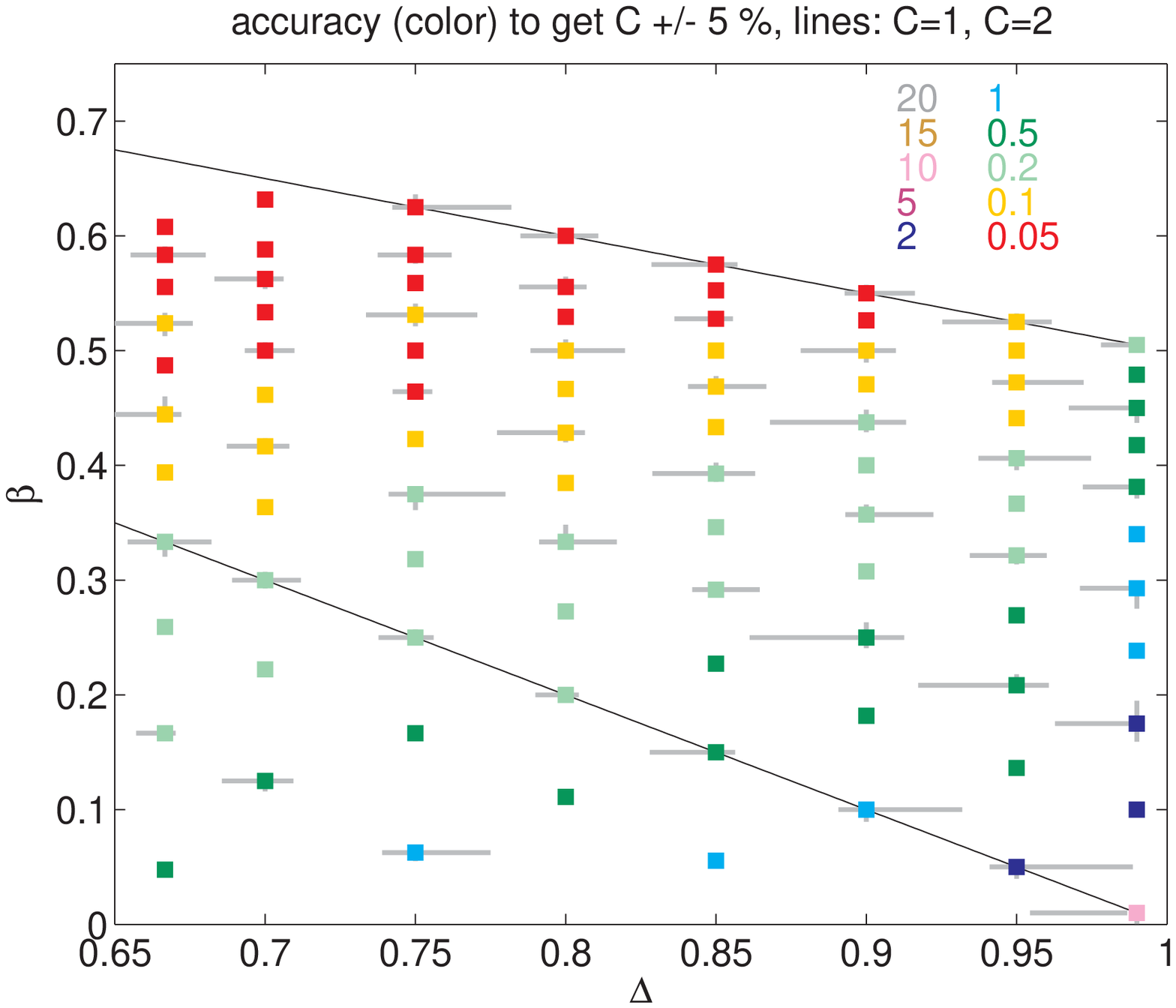}}
\caption{Required accuracy (color coding) of perturbed $Z_{p}$,
  $p=1,.\,.\,.\,,5$, such that for at least 2/3 of the best-fit pairs
  ($\beta_{\mathrm{f}}$, $\Delta_{\mathrm{f}}$) the associated
  $C_{\mathrm{f}}$ are within 10\% (top) or 5\% (bottom)
  of the $C$ associated with the initial, unperturbed ($\beta$,
  $\Delta$) pair. As can be seen, the required accuracy depends
  crucially on the location within the $\beta$-$\Delta$ plane. Gray
  lines indicate the range within which 2/3 of the actual
  $\beta_{\mathrm{f}}$ (vertical direction) and $\Delta_{\mathrm{f}}$
  (horizontal direction) lie. Black lines indicate constant $C=1$
  (lower line) and $C=2$ (upper line). For details see text.}
\label{fig:accu_c_bd_plane}
\end{figure}

Fig.~\ref{fig:accu_c_bd_plane} illustrates the result. Obviously, the
accuracy of the $\tilde{Z}_{p}$ (colored squares, in \%) that is
needed to get at least 2/3 of best-fit $C_{\mathrm{f}}$ to lie within
10\% (top panel) or 5\% (bottom panel) of the exact (initial) $C$
depends on the position within the $\beta$-$\Delta$-plane. In the
lower right parts, $\delta Z_{p}/Z_{p} \ge 1 \mathrm{\%}$ is
sufficient to get 10\% accurate $C_{\mathrm{f}}$, and $\delta
Z_{p}/Z_{p} \ge 0.5 \mathrm{\%}$ yields 5\% accurate $C_{\mathrm{f}}$.
By contrast, in the upper left parts of the plane ($\beta \ge 0.5$)
one needs $\delta Z_{p}/Z_{p} \le 0.1 \mathrm{\%}$ to get 10\%
accurate $C_{\mathrm{f}}$.  Gray lines in
Fig.~\ref{fig:accu_c_bd_plane} (for clarity only shown for a subset of
the colored squares) indicate the range within which at least 2/3 of
the actually fitted $\beta_{\mathrm{f}}$ and $\Delta_{\mathrm{f}}$
lie.  The range is larger for $\Delta$ (horizontal lines) than for
$\beta$ (vertical lines). This is plausible from
Fig.~\ref{fig:dmap_var}, from the area confined by multiple $Z_{p}$
curves. Repeating Fig.~\ref{fig:accu_c_bd_plane} but demanding 10\%
accuracy for $\Delta_{\mathrm{f}}$ instead of $C_{\mathrm{f}}$ yields
a similar pattern in the $\beta$-$\Delta$-plane (not shown), while
demanding 10\% accuracy for $\beta_{\mathrm{f}}$ gives a much more
homogeneous pattern (0.2\% to 0.5\% accuracy for $\tilde{Z}_{p}$, not
shown).

For highly compressible turbulence, best-fit $\beta_{\mathrm{f}}$ and
$\Delta_{\mathrm{f}}$ are expected to lie in the lower part of the
$\beta$-$\Delta$-plane in Fig.~\ref{fig:accu_c_bd_plane}, roughly
$\beta \le 1/3$ and $\Delta \ge 2/3$~\citep{boldyrev:02,
  padoan-et-al:04, pan-et-al:09}. Here, accuracies of 0.5\%, 0.1\%,
and 0.05\% for the $\tilde{Z}_{p}$ translate into accuracies of 10\%,
2\%, and 1\% for $\beta_{\mathrm{f}}$ and $\Delta_{\mathrm{f}}$ (not
shown). Fits of similar quality require much more accurate
$\tilde{Z}_{p}$ in the mildly compressible regime, where $\beta > 1/3$
(upper part of panels in Fig.~\ref{fig:accu_c_bd_plane}) and
ultimately $\beta = 2/3$ in the incompressible limit. We note that in
practical applications, best-fit values may be further improved by
combining, for example, data from different time slices~\citep{pan-et-al:09}.

In summary, a 2\% (1\%) accurate simultaneous fit for
$\beta_{\mathrm{f}}$ and $\Delta_{\mathrm{f}}$ should be possible in
the highly compressible regime if the $Z_{p}$ are 0.1\% (0.05\%)
accurate.  If no satisfying fit is possible for such $Z_{p}$, this may
indicate that the model is not applicable to the turbulence data under
examination.
\section{Discussion}
\label{sec:discussion}
We address three topics. First, can the necessary accuracy for
the $Z_{p}$ be met in practical applications? Second, if we had this type of
accurate simulation data, what could be learned about the hierarchical
structure model and its applicability or non-applicability to driven,
isothermal, supersonic turbulence in a 3D periodic box?  Third, we
want to briefly revisit the frequently used one parameter fits.

We start with the question whether 0.1\% or even 0.05\% accurate
$Z_{p}$ for $p=1,.\,.\,.\,,5$ are achievable, as are needed to get 2\% (1\%)
accurate fits for $\beta$ and $\Delta$. The answer is probably yes, at
least in the context of 3D periodic box simulations.
\citet{schmidt-et-al:08} estimate the accuracy of their $Z_{p}$,
$p=1,.\,.\,.\,,5$, to 1\% (3D box simulations, $1024^{3}$).
\citet{kritsuk-et-al:07} estimate 1\% accuracy or better for absolute
scaling exponents $\zeta_{p}$, $p=1,.\,.\,.\,,3$ (3D box simulations,
$1024^{3}$). Meanwhile, 3D box simulations with $4096^{3}$
exist~\citep[e.g.][]{federrath:13, beresnyak:14}. A first order
estimate suggests the four times better resolution to translate into
four times (first order scheme) or 16 times (second order scheme) more
accurate structure functions, thus accuracies of 0.25\% or even
0.0625\%. Moreover, if the accuracy of the $Z_{p}$ is good enough to
avoid double peaked histograms as in Fig.~\ref{fig:accu_vs_codim},
accuracy may be further enhanced by exploring multiple time slices.
\citet{pan-et-al:09} (3D box simulations, $1024^{3}$) used data from
nine time slices for their two parameter fit. Their work is comparable
although they rely on dissipation rates instead of velocity structure
functions, since the involved scaling exponents ($\tau_{p}$ for the
dissipation rate) are structurally similar according to Kolmogorov's
refined similarity hypothesis~\citep{kolmogorov:62}, $\zeta_{p} = p/3
+ \tau_{p/3}$.  The tests we carried out using ratios of $\tau_{p}$
instead of $\zeta_{p}$ indeed show a similar behavior. From the quality
of their fit and based on our results, we estimate their $\tau_{p}$ to
be about 1\% accurate, which is plausible given their numerical
resolution. A reliable two parameter fit to 3D periodic box data of
highly supersonic turbulence that is based on velocity structure
functions thus appears feasible with today's data.

What could be learned from better simulation data ($4096^{3}$ or
better) and associated, more accurate $Z_{\mathrm{p}}$?  Each
$Z_{\mathrm{p}}$ defines a curve with associated uncertainty in the
$\beta$-$\Delta$ plane, the curves for different $p$ may or may not
intersect to within uncertainties. If they intersect, the model
by~\citet{she-leveque:94} may indeed carry over to highly compressible
turbulence. It is then interesting to see whether the fitted range for
$\Delta$, currently estimated as ($0.67-0.78$)
by~\citet{pan-et-al:09}, remains compatible with $\Delta=2/3$, the
value tacitly assumed in a large body of literature. It is also
interesting to check whether $\beta=1/3$, as theoretically anticipated
by~\citet{boldyrev:02}.  If indeed $(\beta, \Delta) = (1/3, 2/3)$,
results from one parameter fits that quantify the transition to
incompressible turbulence ($\beta=\Delta=2/3$) with decreasing Mach
number likely apply~\citep{padoan-et-al:04}. If $\Delta \ne 2/3$, two
parameter fits for $\beta$ and $\Delta$ would also be needed in the
mildly compressible regime.  However, the analysis in
Sect.~\ref{sec:accu_needed} suggests that these kind of fits are
likely beyond reach of today's computer resources.

The latter of the above cases, where the $Z_{\mathrm{p}}$ curves do not
intersect to within their uncertainty, would imply that the model
by~\citet{she-leveque:94} does not carry over to 3D periodic box
simulations of driven, isothermal, highly compressible turbulence. A
simple reason here could be that theoretical results are based on the
assumption of an infinite Reynolds number, a criterion clearly
violated by numerical simulations. More importantly,
\citet{she-waymire:95} already pointed out that there is no reason why
only one dimension should be associated with the most dissipative
structure. They argued that in such a large portion of space as is
typically analyzed, a variety of most dissipative structures may
co-exist with different co-dimensions.  \citet{hopkins:13} suggests a
slightly different model based on work by~\citet{castaing:96}, which
is more compatible with not strictly log-normal density PDFs as
observed in isothermal supersonic turbulence. Finally, yet other
models exist,~\cite[e.g.  via multifractals][]{macek:11,
  zybin-sirota:13}, as well as other perspectives on the fractal
character of a turbulent medium~\citep[see e.g.][]{kritsuk-et-al:07}.

Lastly, we briefly come back to the one parameter fits that are often
used in the literature. Fixing the value of $\Delta$ by hand greatly
reduces the impact of uncertainties in the $Z_{p}$ on the accuracy of
the estimated best-fit co-dimension $C_{\mathrm{f}}$.
\citet{folini-et-al:14} found 5\% uncertain $Z_{p}$, $p=1,.\,.\,.\,,5$, to
translate into roughly 10\% uncertainty of the $C_{\mathrm{f}}$ for
fixed $\Delta=2/3$.  Fig.~\ref{fig:dmap} offers a qualitative
understanding of this reduced "error propagation", which suggests some
sensitivity to the specific location in the $\beta$-$\Delta$-plane,
and indicates that fixing $C$ or $\beta$ instead of $\Delta$ has a
similar effect.  Fixing $\beta$ seems questionable at first sight
since there is, to our knowledge, little theoretical understanding of
what numerical value $\beta$ might have~\citep[see
e.g.][]{dubrulle:94}. On the other hand, \citet{she-et-al:01}
presented a theoretical framework that allows for an independent
determination of only $\beta$ from the relative scaling exponents
$Z_{p}$ (see also~\citet{hily-blant-et-al:08}, their Appendix A3). One
could thus imagine breaking the two parameter fit for $\beta$ and
$\Delta$ into a two step procedure: first, fix the value of $\beta$,
then use this value and do a one parameter fit for $\Delta$.
Hopefully this type of a two step approach is more robust against
uncertainties in the $Z_{p}$, but this question is beyond the scope of
the current paper and we are unaware of corresponding attempts in the
literature.

\section{Summary and conclusions}
\label{sec:conc}
This study was motivated by the overarching question of whether or not
the random cascade model~\citep{she-leveque:94, dubrulle:94,
  she-waymire:95, boldyrev:02} applies to simulation data of highly
compressible isothermal turbulence and, if so, with what parameter
values for $\beta$ and $\Delta$. If applicable, the model offers a
theoretical link between observable properties of the turbulence,
namely ratios $Z_{p}$ of scaling exponents of the structure functions,
and non-observable turbulence characteristics, for example the
dimension $D$ of the most dissipative structures. To date,
applicability of the model is assumed in much of the literature with
$\Delta=2/3$, a value just marginally compatible with simulation-based
best estimates~\citep{pan-et-al:09}: $\Delta=0.71$ with an uncertainty
range $\Delta \in (0.67, 0.78)$.

We examine how uncertainties in the $Z_{p}$ translate into
uncertainties of best-fit $\beta$-$\Delta$-pairs and discuss what
best-fits, consequently, seem achievable with today's computer
resources. A Monte Carlo approach is used to mimic actual simulation
data. The results can be summarized in six main points.

\begin{itemize}
\item{Simultaneous fitting of $\beta$ and $\Delta$ to sets of
    substantially (5\%) perturbed (uncertain) $Z_{p}$ yields a "double
    peaked ridge" of best-fit values in the $\beta$-$\Delta$ plane.
    None of the two peaks is co-located with the initial $(\beta,
    \Delta)$ pair.}

\item{The highest and lowest order $p$ are particularly relevant for
    simultaneous fitting of $\beta$ and $\Delta$. A somewhat optimal
    choice is $p=1,.\,.\,.\,,5$. Yet higher order structure functions add
    comparatively little to the quality of the fit, while they tend to
    be afflicted with larger uncertainties in real applications.}

\item{A simultaneous, 2\% (1\%) accurate fit of $\beta$ and $\Delta$
    should be possible if the $Z_{p}$, $p=1,.\,.\,.\,,5$, are~0.1\% (0.05\%)
    accurate and if the (yet to be determined) value of $\beta$ is
    about 1/3 or less.}

\item{Applicability of the model thus may be best tested in the highly
    compressible regime, where $\beta \approx 1/3$ is expected, and
    not in the mildly compressible regime where $\beta$ ultimately
    must approach its incompressible value of 2/3.}

\item{We argue that today's computer resources likely allow to reach
    this accuracy. Existing simulations of
    $4096^{3}$~\citep{federrath:13, beresnyak:14} probably allow for
    at least 2\%, possibly 1\% accurate estimates of $\beta$ and
    $\Delta$.}

\item{Should the ambiguity in the determination of $\beta$ and
    $\Delta$ persist despite such highly accurate $Z_{p}$, this may
    indicate that the notion of~\citet{she-waymire:95} ($\beta$ and
    $\Delta$ take a continuum of values) or~\citet{hopkins:13} (a
    different model for the statistics of the inertial range) is
    correct or that yet a different turbulence model is needed in this
    regime.}
\end{itemize}

While the authors lack the computational resources to really test the
estimates presented here, this study may encourage other groups to
analyze their data in the light of this study.
\begin{acknowledgements}
  RW and DF acknowledge support from the French National Program for
  High Energies PNHE.  We acknowledge support from the P\^{o}le
  Scientifique de Mod\'{e}lisation Num\'{e}rique (PSMN), from the
  Grand Equipement National de Calcul Intensif (GENCI), project number
  x2014046960, and from the European Research Council through grant ERC-AdG
  No. 320478-TOFU.

\end{acknowledgements}
%
%
%
%
%
%
%
\bibliographystyle{apj} 
\bibliography{BigAstroBib} 
%
%
%
%
%
%
%
%
%
%

\end{document}